%% file: draft_revised_final.tex
\documentclass[notitlepage,aps,superscriptaddress,showpacs,showkeys,nofootinbib,floatfix]{revtex4-1}

\usepackage{amsmath,amssymb,amscd,,slashed,multirow,amsfonts,latexsym}
\usepackage{listings}
\usepackage{dsfont}
\usepackage{slashed}
\usepackage{xcolor}

\usepackage{graphicx,wrapfig,epsfig}
\newsavebox{\imagebox}
\usepackage{subcaption}
\usepackage{ragged2e}
\DeclareCaptionJustification{justified}{\justifying}
\usepackage{epstopdf}
\usepackage[breaklinks,hidelinks]{hyperref}

\newcommand{\bea}{\begin{eqnarray}}
\newcommand{\beal}[1]{\begin{eqnarray}\label{#1}}
\newcommand{\eea}{\end{eqnarray}}
\newcommand{\be}{\begin{equation}}
\newcommand{\ee}{\end{equation}}
\newcommand{\beq}{\begin{equation}}
\newcommand{\beql}[1]{\begin{equation}\label{#1}}
\newcommand{\enq}{\end{equation}}
\newcommand{\betab}{\begin{table}}
\newcommand{\entab}{\end{table}}

\newcommand{\mc}{\mathcal}
\newcommand{\f}{\frac}
\newcommand{\np}{{\nu^\prime}}
\newcommand{\tp}{{t^\prime}}
\def\astat{\alpha_{E1}}
\def\bstat{\beta_{M1}}

\newcommand{\benum}{\begin{enumerate}}
\newcommand{\enum}{\end{enumerate}}
\newcommand{\beitem}{\begin{itemize}}
\newcommand{\enitem}{\end{itemize}}
\newcommand{\eqr}[1]{Eq.~(\ref{#1})}

\newcommand{\tref}[1]{Table~\ref{#1}}

\newcommand{\fr}[1]{Fig.~\ref{#1}}
\newcommand{\frtwo}[2]{Figs.~\ref{#1} and \ref{#2}}

\newcommand{\cref}[1]{Ch.~\ref{#1}}
\newcommand{\secref}[1]{Sec.~\ref{#1}}
\newcommand{\apref}[1]{App.~\ref{#1}}
\newcommand{\ie}{i.e.,~}
\newcommand{\nn}{\nonumber\\}

\newlength\savedwidth

\usepackage[normalem]{ulem}

\definecolor{olive}{HTML}{668000}
\definecolor{lightolive}{HTML}{CCFF00}
\definecolor{darkolive}{HTML}{446600}
\definecolor{myorange}{HTML}{FF9900}
\definecolor{myblue}{HTML}{8080FF}
\definecolor{mygreen}{HTML}{008000}
\definecolor{purple1}{HTML}{BB99FF}
\definecolor{joker}{HTML}{990099}
\definecolor{fucsia}{HTML}{FF3399}
\definecolor{myred}{HTML}{FF0000}

\def\oxl{{\rotatebox[origin=c]{90}{\large{\mypentagon{myorange}}}}}
\def\hym{\large{$\textcolor{blue}\bullet$}}
\def\gold{{\rotatebox[origin=c]{180}{\large{\mytriangle{olive}}}}}
\def\ber{{\rotatebox[origin=c]{0}{\large{\mytriangle{purple1}}}}}
\def\pugh{{\rotatebox[origin=c]{90}{\large{\mypentagon{darkolive}}}}}
\def\baranov{{\rotatebox[origin=c]{180}{\large{\mytriangle{green}}}}}
\def\baranovbis{{\rotatebox[origin=c]{180}{\large{\myemptytriangle{green}}}}}
\def\fed{{\rotatebox[origin=c]{0}{\large{\mytriangle{myblue}}}}}
\def\zie{{\rotatebox[origin=c]{45}{\large{\mysq{joker}}}}}
\def\hal{{\rotatebox[origin=c]{0}{\large{\mysq{black}}}}}
\def\mcg{{\rotatebox[origin=c]{0}{\large{\mysq{myred}}}}}
\def\mcgbis{{\rotatebox[origin=c]{0}{\large{\myemptysq{myred}}}}}
\def\taps{{\rotatebox[origin=c]{45}{\large{\mysq{fucsia}}}}}

\draft

\allowdisplaybreaks

\usepackage{tikz}

\newcommand{\mypentagon}[1]{
\begin{tikzpicture}
\fill[opacity=1,#1] (1ex, 0.5257311121191336ex) -- (0.6180339887498949ex,1.0514622242382672ex) -- (0, 0.8506508083520399ex) -- (0, 0.2008114158862273ex) -- (0.6180339887498949ex, 0) -- (1ex, 0.5257311121191336ex);
\end{tikzpicture}
}
\newcommand{\mytriangle}[1]{
\begin{tikzpicture}
\fill[opacity=1,#1] (-0.5ex, 0) -- (0.5ex, 0) -- (0.ex, 0.866025403784ex) -- (-0.5 ex,0);
\end{tikzpicture}
}

\newcommand{\myemptytriangle}[1]{
\begin{tikzpicture}
\draw[#1] (-0.5ex, 0) -- (0.5ex, 0) -- (0.ex, 0.866025403784ex) -- (-0.5 ex,0);
\end{tikzpicture}
}

\newcommand{\mysq}[1]{
\begin{tikzpicture}
\fill[opacity=1,#1] (-0.5ex, 0) -- (0.5ex, 0) --  (0.5ex, 1ex) -- (-0.5ex,1ex) -- (-0.5ex, 0);
\end{tikzpicture}
}

\newcommand{\myemptysq}[1]{
\begin{tikzpicture}
\draw[#1] (-0.5ex, 0) -- (0.5ex, 0) --  (0.5ex, 1ex) -- (-0.5ex,1ex) -- (-0.5ex, 0);
\end{tikzpicture}
}

\usepackage{wasysym}
\usepackage{marvosym}

\begin{document}

\title{Proton scalar dipole polarizabilities from real Compton scattering data, using fixed-t subtracted dispersion relations and the bootstrap method}
\author{B.~Pasquini}
\affiliation{Dipartimento di Fisica,
Universit\`a degli Studi di Pavia, 27100 Pavia, Italy}
\affiliation{Istituto Nazionale di Fisica Nucleare,
Sezione di Pavia, 27100 Pavia, Italy}
\author{P. Pedroni}
\affiliation{Istituto Nazionale di Fisica Nucleare,
Sezione di Pavia, 27100 Pavia, Italy}
\author{S. Sconfietti}
\affiliation{Dipartimento di Fisica,
Universit\`a degli Studi di Pavia, 27100 Pavia, Italy}
\affiliation{Istituto Nazionale di Fisica Nucleare,
Sezione di Pavia, 27100 Pavia, Italy}

\begin{abstract}
We perform a fit of the real Compton scattering (RCS) data below pion-production threshold to extract the  electric ($\alpha_{E1}$) and magnetic ($\beta_{M1}$) static scalar dipole polarizabilities of the proton,  using fixed-$t$ subtracted dispersion relations and a bootstrap-based fitting technique. 
The bootstrap method provides a convenient tool to  include the effects of the systematic errors on the best values of $\astat$ and $\bstat$
and
to propagate the statistical errors of the  model parameters fixed by other measurements. We also implement various statistical tests to investigate the consistency of the
available RCS data sets below pion-production threshold and we conclude that there are not strong motivations to exclude any data point   from the global set.
Our analysis yields $\astat = (12.03^{+0.48}_{-0.54})\times 10^{-4} \text{fm}^3$ and $\bstat = (1.77^{+0.52}_{-0.54})\times 10^{-4} \text{fm}^3$, with p-value $= 12\%$.
\end{abstract} 

\maketitle

\section{Introduction}
The electric and magnetic static scalar dipole polarizabilities, $\alpha_{E1}$ and $\beta_{M1}$, respectively, are fundamental structure constants of the proton that can be accessed via real Compton scattering (RCS).
In the low-energy expansion of the Compton amplitude, they correspond to the leading-order contributions beyond the structure independent terms that describe  the scattering process 
as if the proton were a pointlike particle with anomalous magnetic moment. When approaching the pion-production threshold, also higher-order terms start competing  with the scalar dipole polarizabilities. Therefore, one has to resort  
 to reliable theoretical frameworks for extracting the scalar dipole polarizabilities from experimental data.
The most accredited theories, which have been used sofar, are fixed-$t$ dispersion relations (DRs), in the unsubtracted~\cite{Lvov:1996rmi,Babusci:1998ww,Schumacher:2005an} and subtracted~\cite{Drechsel:1999rf,Holstein:1999uu,Pasquini:2007hf,Drechsel:2002ar,Pasquini:2018wbl} formalism, and chiral perturbation theory ($\chi$PT) with explicit nucleons and Delta's, in the variant of heavy-baryon
$\chi$PT (HB$\chi$PT)~\cite{Bernard:1995dp,Beane:2002wn,McGovern:2012ew} and manifestly covariant~\cite{Lensky:2015awa,Lensky:2009uv} $\chi$PT (B$\chi$PT)~\footnote{We refer to~\cite{Kondratyuk:2001qu,Gasparyan:2011yw} for other theoretical predictions of the scalar and spin polarizabilities not fitted to experimental data.}.
Based on these theoretical frameworks, extractions of the scalar dipole polarizabilities have been obtained by fitting different data sets for the unpolarized RCS cross section, and adopting a statistical approach based on the conventional $\chi^2$-minimization procedure.
Recently, a new statistical method has successfully been applied in Ref.~\cite{Pasquini:2017ehj} to analyze RCS data at low energies and extract values for the energy-dependent scalar dipole dynamical polarizabilities~\cite{Griesshammer:2001uw,Hildebrandt:2003fm}.
The method is based on the parametric-bootstrap technique, and  it is adopted in this work to extract the scalar dipole  static polarizabilities, using the updated  version of  fixed-$t$ subtracted DRs formalism~\cite{Pasquini:2018wbl} as theoretical framework.
  Although the bootstrap method is rarely used in nuclear physics~\cite{NavarroPerez:2018qzd,Pasquini:2017ehj,Perez:2014jsa,Nieves:1999zb,Bertsch:2017ilb}, it has high potential and advantages~\cite{Pastore:2018xuu}. 
In particular, we will show that
it allows us to   include the systematic errors in the data analysis in a straightforward way and to efficiently reconstruct the probability distributions of the fitted parameters.
We will also pay a special attention to discuss the available  sets of RCS data below pion-production threshold. Following recent discussions about the possible presence of outliers in the available data sets~\cite{Krupina:2017pgr,Griesshammer:2012we}, we perform
several tests to judge the data-set consistency.

The  manuscript is organized as follows. In \secref{sec:theory}, we briefly summarize the theoretical framework of fixed-$t$ subtracted DRs.
In \secref{sec:fit1}, we describe the main features of  the parametric-bootstrap technique, which is applied in 
 \secref{sec:fit2} to our specific case to fit  $\alpha_{E1}$ and $\beta_{M1}$. We perform the fit in different conditions, i.e., switching on/off the effects of the systematic errors, using the constraint of the Baldin's sum rule for the polarizability sum and including the backward spin polarizability $\gamma_\pi$ as additional fit parameter. 
The consistency of the data set is discussed in~\secref{sec:discussion}, 
where we perform different statistical tests to identify the possible presence of outliers.
The results of our analysis are summarized in~\secref{sec:summary}, in comparison with available extractions of the scalar dipole polarizabilities. Our conclusions are drawn in Sec.~\ref{sec:conclusions}.
In \apref{sec:data}, we give the complete list of the existing data sets of RCS below pion-production threshold, and in~\apref{sec:correlations}
we discuss the values of the correlations among the fit parameters in all the different conditions discussed in this work.

\section{Theoretical framework}
\label{sec:theory}
We consider RCS off the proton, i.e. $\gamma(q)+P(p)\rightarrow \gamma(q^\prime)+P(p^\prime)$,
where the variables in brackets denote the four-momenta of the participating particles.
The familiar Mandelstam variables are $s=(p+q)^2$, $u=(q-p^\prime)^2$ and $t=(q-q^\prime)^2$, and are constrained by $s+u+t=2M^2$, with $M$ the proton mass.
The RCS amplitude can be described in terms of 6 Lorentz invariant functions $A_i(\nu,t)$, which depend on the crossing-symmetric variable $\nu=(s-u)/4M$ and $t$.
They are free of kinematical singularities and constraints, and because of the crossing symmetry they obey the relation $A_i(\nu, t) = A_i(-\nu, t)$.
Assuming analyticity, they satisfy the following fixed-$t$ subtracted DRs (with the subtraction point at $\nu=0$)~\cite{Drechsel:1999rf,Pasquini:2007hf}
\beql{eq:drs2}
\mathrm{Re}[A_i(\nu,t)] = A_i^B(\nu,t) +\left[A_i(0,t)- A_i^B(0,t)\right]
+\f{2}{\pi}\nu^2\mc P \int_{\nu_0}^{+\infty}d\np\f{\mathrm{Im}_s[A_i(\nu^\prime,t)]}{\np(\np^2-\nu^2)},
	\enq
where $\nu_0$ is the pion-production threshold, and $A_i^B(\nu,t)$ is the Born term, corresponding to the pole diagrams
involving a single nucleon exchanged in $s$- or $u$-channel
and $\gamma N N$ vertices taken in the on-shell regime.
In Eq.~\eqref{eq:drs2}, the subtraction functions $\left[A_i(0,t)- A_i^B(0,t)\right]
$ can be determined	by once-subtracted DRs in the $t$ channel:
\beal{eq:drs3}
A_i(0,t)- A_i^B(0,t) &=&\left[A_i(0,0)- A_i^B(0,0)\right] +\left[A_i^{t-pole}(0,t)- A_i^{t-pole}(0,t)\right] \nn
&+&  \f{t}{\pi} \int_{4m_\pi^2}^{+\infty}d\tp\f{\mathrm{Im}_t[A_i(0,\tp)]}{\tp(\tp-t)} + \f{t}{\pi} \int_{-\infty}^{-2m_\pi^2-4Mm_\pi}d\tp\f{\mathrm{Im}_t[A_i(0,\tp)]}{\tp(\tp-t)},
	\eea
	where $A_i^{t-pole}(0,t)$ represents the contribution of the poles in the $t$ channel, that amounts to the $\pi^0$-pole contribution to the $A_2$ amplitude.
	The subtraction constants $a_i \equiv \left[A_i(0,0)- A_i^B(0,0)\right]$ are directly related to the scalar dipole and leading-order spin polarizabilities, i.e.
	\begin{equation}
	\begin{split}
	\alpha _{{E1}}&=\f{ -a_1-a_3-a_6}{4\pi},\\
	\gamma_{{E1E1}}&=
  \frac{a_2-a_4+2 a_5+a_6}{8\pi M_N},\\
\gamma_{{M1E2}}&=
  \frac{-a_2-a_4-a_6}{8\pi M_N},\\
  \end{split}
  \qquad
  \begin{split}
  \beta_{{M1}}&= \f{a_1-a_3-a_6}{4\pi},\\
\gamma_{{M1M1}}&=
  \frac{-a_2-a_4-2 a_5+a_6}{8\pi M_N},\\
  \gamma_{{E1M2}}&=
  \frac{a_2-a_4-a_6}{8\pi M_N},
\end{split}
	\end{equation}
with the combination 
	\beq
\gamma_0 \equiv -\gamma_{E1E1} - \gamma_{M1M1} -\gamma_{E1M2} - \gamma_{M1E2}, \quad \gamma_\pi \equiv -\gamma_{E1E1} + \gamma_{M1M1} -\gamma_{E1M2} + \gamma_{M1E2}
	\enq
defining the forward ($\gamma_0$) and backward ($\gamma_\pi$) spin polarizabilities.
We will consider $\{\gamma_{E1E1},\gamma_{M1M1},\gamma_0,\gamma_\pi\}$ as independent set of spin polarizabilities.

In the actual calculation, the $s$-channel imaginary parts in Eq.~\eqref{eq:drs2} are evaluated using the unitarity relation, taking into account the contribution of the 
$\pi N$ intermediate states from the latest version of the MAID pion-photoproduction amplitudes~\cite{Drechsel:2007if} and approximating the contribution from multipion intermediate channels by 
the inelastic decay channels of the $\pi N$ resonances, as detailed in Ref.~\cite{Drechsel:1999rf}.
It was found, however, that in the subtracted dispersion relation formalism, the sensitivity to the multipion channels is very small and that subtracted dispersion
relations are
essentially saturated at $\nu\approx 0.4$ GeV.
Furthermore, the $t$-channel imaginary parts in Eq.~\eqref{eq:drs3} are calculated using the $\gamma\gamma\rightarrow \pi \pi \rightarrow N\bar{N}$ channel as input for the positive-$t$ cut, while the negative-$t$ cut is strongly suppressed for low values of $t$. The last one can be approximated by the contributions of $\Delta$-resonance and non-resonant $\pi N$ intermediate states in the $s$-channel, which are then
extrapolated into the unphysical region at $\nu=0$  
by analytical continuation.
For more detail in the implementation of the unitarity relations, we refer to the original work~\cite{Drechsel:1999rf}.
Having determined the contributions of the $s$- and $t$-channel integrals, the only remaining unknown are the subtraction constants, i.e. the leading-order static polarizabilities.
In principle, all the six leading static polarizabilities can be used as free fit parameters to the Compton observables.
However, a simultaneous fit of all them is not feasible at the moment, because of the limited statistics of the available RCS data.
In the following, we will limit ourselves to the 
data sets for unpolarized RCS below pion-production threshold, and consider different variants of fits for two sets of parameters, i.e. $\{\alpha_{E1},\beta_{M1}\}$ or 
$\{\alpha_{E1},\beta_{M1},\gamma_\pi\}$. 
The remaining constants which do not enter the fit are fixed as described in \secref{sec:fit2}.

\section{The fitting method}
\label{sec:fit1}

We consider a generic problem, where we have a model prediction $T(p)$  for an observable,
which depends on a set $p$ of parameters, and we want
to find the optimal set $\hat{p}$   that better reproduces the available experimental data.
We adopt an algorithm
based on the parametric bootstrap technique~\cite{Davidson-Hinkley}, i.e., $N$ Monte Carlo replicas of experimental data are produced and a
fit of the set $p$ is performed to every bootstrapped data sample. 
After every cycle $j$, the best values  $\hat{p}_j$ are stored, to obtain  $N$ outcomes of the (unknown)
probability distribution of $p$.  

In our case  we assume that:
\benum
\item every data point is Gaussian distributed with a mean equal to the measured value and a standard deviation given by the experimental (statistical) error;
\item
  data points are affected by systematic errors given by different rescaling factors of the data in each subset;  
\item
 when not explicitly stated otherwise by the experimental groups,
  every source of systematic error follows an uniform
  distribution and the published value gives the full estimated interval. 
If there are more sources, we take the product of such random uniform variables;
\item the sample in every data subset is independent from the other subsets.
  \enum
This sampling method can then be written  as
	\beql{eq:chiboot}
\mc S_{ij} = (1 + \delta_{ij})( E_{i}+\gamma_{ij}\sigma_i),
\enq
where $S_{ij}$ is a generic bootstrapped point with
the index $i$ running over the number of data point ($n_{data}$) and $j$
running over the number of replicas ($N$).
$E_i$ is the  experimental point having an uncertainty $\sigma_i$, $\gamma_{ij}$ is
the Gaussian normal variable needed for the statistical sampling and $\delta_{ij}$ is a box distributed variable that quantifies the effect of the systematic uncertainties for each data subset independently.
Considering a generic subset, labeled with $k$ ($k$ runs from 1 to the number of the different data subsets
$n_{set}$) and composed of $n_k$ data points,
we take $\delta_{ij} \in \mc U[-\Delta_k,\Delta_k]$ for $i=1,\dots, n_k$, where $\pm\Delta_k$ is the published systematic
error and $\sum_{k=1}^{n_{set}} n_k = n_{data}$.
If there are $n_s$ different and independent sources of systematic uncertainties, $\delta_{ij}$ is the product
of all the $n_s$ box distributed variables, \ie $\delta_{ij}=\prod_{f=1}^{n_s}\mc U[-\Delta_f,\Delta_f]$.
The systematic sources can be easily excluded from this procedure by just imposing $\delta_{ij}\equiv 0$
in \eqr{eq:chiboot}.

The minimization function at the $j^{th}$ iteration is given by
	\beql{eq:chiboot2}
\chi_{b,j}^2 = \sum_{i=1}^{n_{data}} \left(\f{\mc S_{ij}-T_{i}(p)}{\sigma_{ij}}\right)^2,
\enq
where
  \beql{eq:chiboot2b}
  \sigma_{ij} = (1 + \delta_{ij})\sigma_i.
  \enq
The minimum in the parameter space can be defined as
\beql{eq:chiboot4}
{\hat\chi_{b,j}}^2 = \sum_{i=1}^{n_{data}} \left(\f{\mc S_{ij}-T_i(\hat p_j)}{\sigma_{ij}}\right)^2,
	\enq
where $\hat p_j$ are the best values of the 
fit parameters $p$ at the $j^{th}$ bootstrap cycle.

  Repeating this minimization for $N$ cycles,
  the empirical distribution $\mc P(\hat p_j)$ of the $\hat p_j$ random variables
  gives an estimate of the true
  probability distribution $\mc P(p)$ that includes the propagation of both statistical and systematic errors of the experimental
  data. The best value and the standard deviation of $p$ can be then simply obtained as:
  	\beql{eq:chiboot5}
        \hat p \equiv \frac{1}{N} \sum_{j=1}^N \hat p_j , \qquad \sigma_p \equiv
        \left(
        \f{1}{N-1} \sum_{j=1}^N (\hat p_j - \hat p )^2 
        \right)^{1/2}.
        \enq
        
The goodness of  this fit procedure can be estimated in the same way as in the standard case, using the
value $\hat \chi^2$ of the so-called $\chi^2$-variable, defined as~\footnote{
The link between ${\hat\chi_{b,j}}^2$ and $\hat\chi^2$ can be found in Ref.~\cite{stat_paper}.}:

\beql{eq:chiboot3}
\hat \chi^2 = \sum_{i=1}^{n_{data}} \left(\f{E_i-T_i(\hat p )}{\sigma_i}\right)^2.
\enq

It is worthwhile to notice here that $\hat \chi^2$ is distributed according to the  $\chi^2$ distribution
only when $\delta_{ij}=0$, i.e. when all the $E_i$ are independent random gaussian variables.


  Within the bootstrap framework, it is also possible to evaluate the expected theoretical probability
  distribution associated to ${\hat\chi}^2$ by 
  replacing $\mc S_{ij}$ in \eqr{eq:chiboot2} with
    
    \beql{eq:thboot1}
    \mc M_{ij} = (1 + \delta_{ij})( T_i(\hat p)+\gamma_{ij}\sigma_i),
    \enq
and by finding, at each bootstrap cycle, the minimum value $ \hat{\chi}_{th,j}^2$ of the following function 
\beql{eq:thboot2}
  \chi_{th,j}^2 = \sum_{i=1}^{n_{data}} \left(\f{\mc M_{ij}-T_i(p_j)}{\sigma_{ij}}\right)^2 .
\enq

After $N$ bootstrap iterations,  we are able to empirically reconstruct the probability distribution
$\mc P(\chi_{th}^2)$
 and then to evaluate the final p-value associated to the fit.

It can be easily demonstrated (see~\cite{stat_paper}) that, when
 $\delta_{ij}=0$ in \eqr{eq:thboot1},
$\mc P(\chi_{th}^2)$ coincides with the  $\chi^2$ distribution, as expected.
In any case, we
stress that the bootstrap method allows us to  obtain a p-value for $\hat\chi^2$ 
directly from the evaluated $\mc P(\chi_{th}^2)$ distribution,
  also when systematic errors
are taken into account in the fit procedure.

\subsection{Uncertainties on  additional model parameters}
\label{sec:fix_par}

In the most generic situation, the model $T$ may depend on an additional set
of  parameters $f$ besides the fit parameters $p$, \ie $T \equiv T(p,f)$. The ${\chi_{b,j}}^2$ variable of \eqr{eq:chiboot2} is consequently modified as
	\beql{eq:error1}
{\chi_{b,j}}^2 = \sum_{i=1}^{n_{data}} \left(\f{\mc S_{ij}-T_i(p,f)}{\sigma_{ij}}\right)^2.
	\enq

Suppose the values of the parameters $f$ are derived from experimental data and are known
within an experimental uncertainty $\sigma_{f}$.
Within the bootstrap framework,
we can easily evaluate  how the uncertainties $\sigma_{f}$
affect the values of the fit parameters $\hat p$,
without using the error-propagation procedure that would require performing numerical derivatives
$\partial T/\partial f$.
At each $j^{th}$ bootstrap cycle, we can sample the value  $f_{j}$ of the  model parameters   from
their known probability distribution, which in the following 
will be considered to be a Gaussian defined as $\mc G[f,\sigma^2_{f}]$.
Then, 
we can repeat the procedure described above by
  replacing $T_i(\hat p_j)$ with $T_i(\hat p_j,f_j)$ in \eqr{eq:chiboot4}, 
  and evaluate all the relevant fit parameters.

	\section{Fit to RCS data}
		\label{sec:fit2}
In this section, we apply the fitting method introduced in \secref{sec:fit1} to analyze available RCS data below pion-production threshold.
We use fixed-$t$ subtracted DRs for the model predictions, which contain the leading-order static polarizabilities as free parameters, as explained in \secref{sec:theory}.
We discuss two data sets, corresponding to the FULL and TAPS data sets, as described in \apref{sec:data}.
Furthermore, we consider different fit conditions, switching on/off the systematic errors and using two sets of free parameters:
 {\it i)} the scalar dipole polarizabilities, with and without the constraint of the Baldin's sum rule for the polarizability sum $\alpha_{E1}+\beta_{M1}$, and {\it ii)}
the scalar dipole polarizabilities constrained by the Baldin's sum rule  along with the backward spin polarizability $\gamma_\pi$. 
For the Baldin's sum rule, we use the weighted
 average over the available evaluations reported in Ref.~\cite{Hagelstein:2015egb}, which coincides also with the value used in the fit of Refs.~\cite{OlmosdeLeon:2001zn,McGovern:2012ew,Martel:2014pba}, \ie
$\astat+\bstat =13.8\pm 0.4$.
The remaining parameters of fixed-$t$ subtracted DRs are fixed to the experimental values extracted from double polarization RCS~\cite{Martel:2014pba}, i.e. $\gamma_{E1E1}=-3.5\pm 1.2$ and $\gamma_{M1M1}=3.16\pm 0.85$, 
and from the GDH experiments~\cite{Ahrens:2001qt,Dutz:2003mm}, i.e. 
$\gamma_0=-1.01\pm 0.08\pm 0.10$~\footnote{This value is consistent with the fitting conditions adopted for the extraction of the spin polarizability in Ref.~\cite{Martel:2014pba}.
We note that recent reevaluations~\cite{Pasquini:2010zr,Gryniuk:2016gnm} of $\gamma_0$ give a slightly smaller central values, with uncertainties consistent with the value used in Ref.~\cite{Martel:2014pba}.}.
When the backward spin polarizability is not used as fit parameter, we fixed it to  the weighted average of the values extracted at MAMI~\cite{Schumacher:2005an}, i.e. $\gamma_\pi=-8.0\pm 1.8$.
Here and in the following, we used the standard convention to exclude the $t$-channel $\pi^0$-pole contribution from the spin polarizabilities. These contributions amount to
 ${\gamma_\pi}^{\pi^0-\text{pole}}=-46.7$~\cite{Pasquini:2007hf}, 
  $\gamma_{M1M1}^{\pi^0-\text{pole}}=-\gamma_{E1E1}^{\pi^0-\text{pole}}=
  \tfrac{1}{4}
  \gamma_\pi^{\pi^0-\text{pole}}
  $, while they vanish in the case of the forward spin polarizability.
 Finally,
for each fitting configuration, we discuss the probability distributions of the fitted parameters and the p-values of the $\hat\chi^2$ variable.
Here and in the following,
 we use  the 
 units  of $10^{-4}$~fm$^3$
for the scalar dipole polarizabilities
 and $10^{-4}$~fm$^4$ for the spin polarizabilities.

 \subsection{ Handling  the experimental and model errors} 
 \label{sec:errprop}
 We apply the  method described above using $N=10000$ bootstrap replicas.
Within this framework and following the method outlined in \secref{sec:fix_par},
 we take into account
the uncertainties of the model parameters on the values of the polarizabilities not treated as free parameters in the fit procedure.
In particular, we take $\gamma_0 \in \mc G[-1.01,0.13^2]$\footnote{The uncertainty value $0.13^2$ is the sum of the squares of the statistical and systematic errors.}, $\gamma_{E1E1} \in \mc G[-3.5,1.2^2]$ and $\gamma_{M1M1} \in \mc G[3.16,0.85^2]$. 
When keeping fixed the backward spin polarizability, we propagate the error of $\gamma_\pi$ 
using $\gamma_\pi\in \mc G[8.0,1.8^2]$. Furthermore, the Baldin's sum rule constraint is implemented using $\astat+\bstat \in \mc G[13.8,0.4^2]$.
The uncertainties on the fitted $\astat$ and $\bstat$ thus automatically include the propagation of the errors of the spin polarizabilities and the Baldin's sum rule.
The statistical and systematic uncertainties of the experimental data are taken into account as described
in \secref{sec:fit1}, except for the TAPS data points~\cite{OlmosdeLeon:2001zn}.
		As discussed in Ref.~\cite{Griesshammer:2012we}, they are affected by a $5\%$ point-to-point systematic error, and, accordingly, the statistical error of each point is modified as follows
		\beq
\sigma_{i,TAPS} \rightarrow \left[\sigma_{i,TAPS}^2 + \left(\f{5}{100}E_{i,TAPS}\right)^2\right]^{1/2}.\label{eq:taps-systematic}
	\enq

\subsection{Results}
\label{subsect:results}
\begin{figure}[h]
\includegraphics[scale=1.]{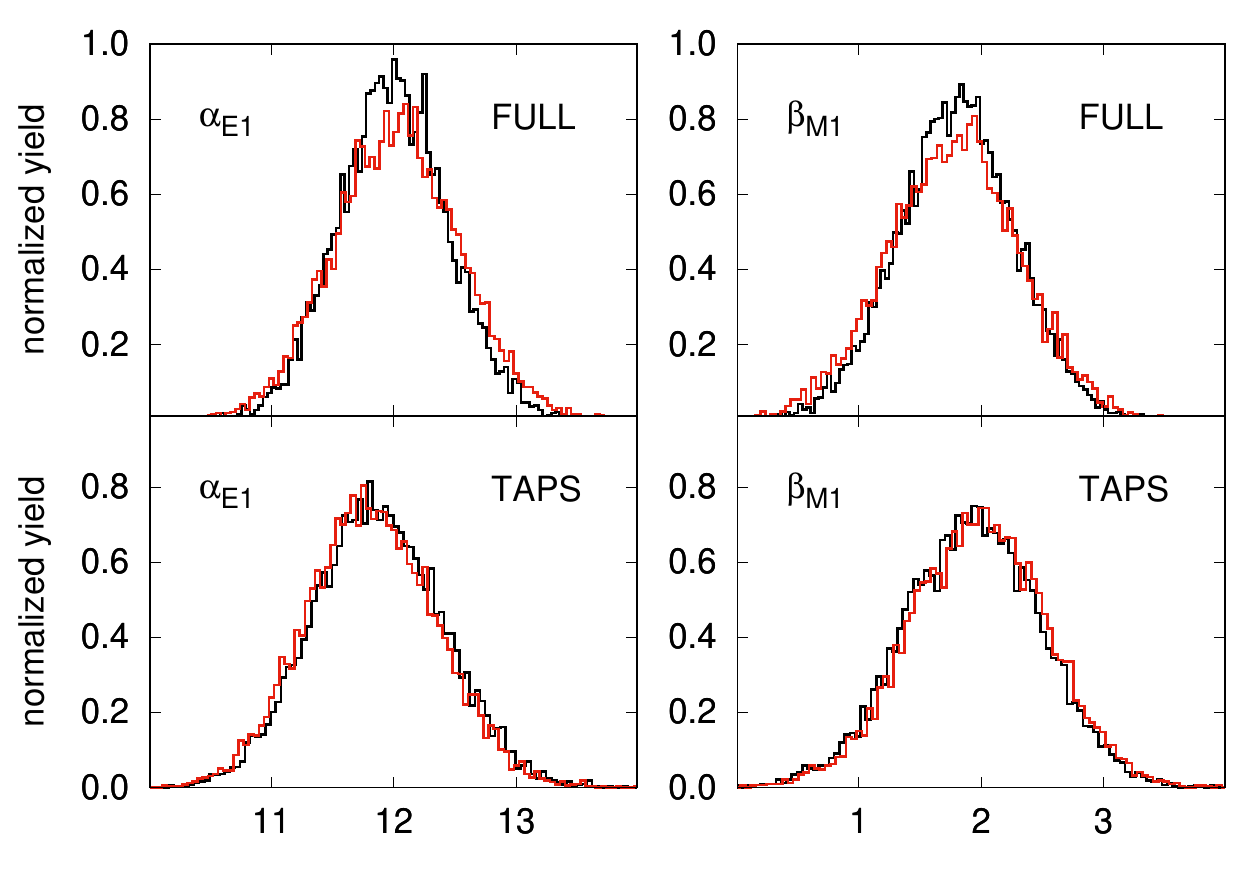}
\caption{Probability distributions of the fitted scalar dipole static polarizabilities $\astat$ (left panels) and $\bstat$ (right panels) in the Fit 1 (black curve) and Fit $1^\prime$ (red curve) conditions.
The results are obtained using the FULL data set (upper panels) and the TAPS data set (lower panels).\label{Fig1}}
	\end{figure}

\begin{figure}[h!]
\includegraphics[scale=1.]{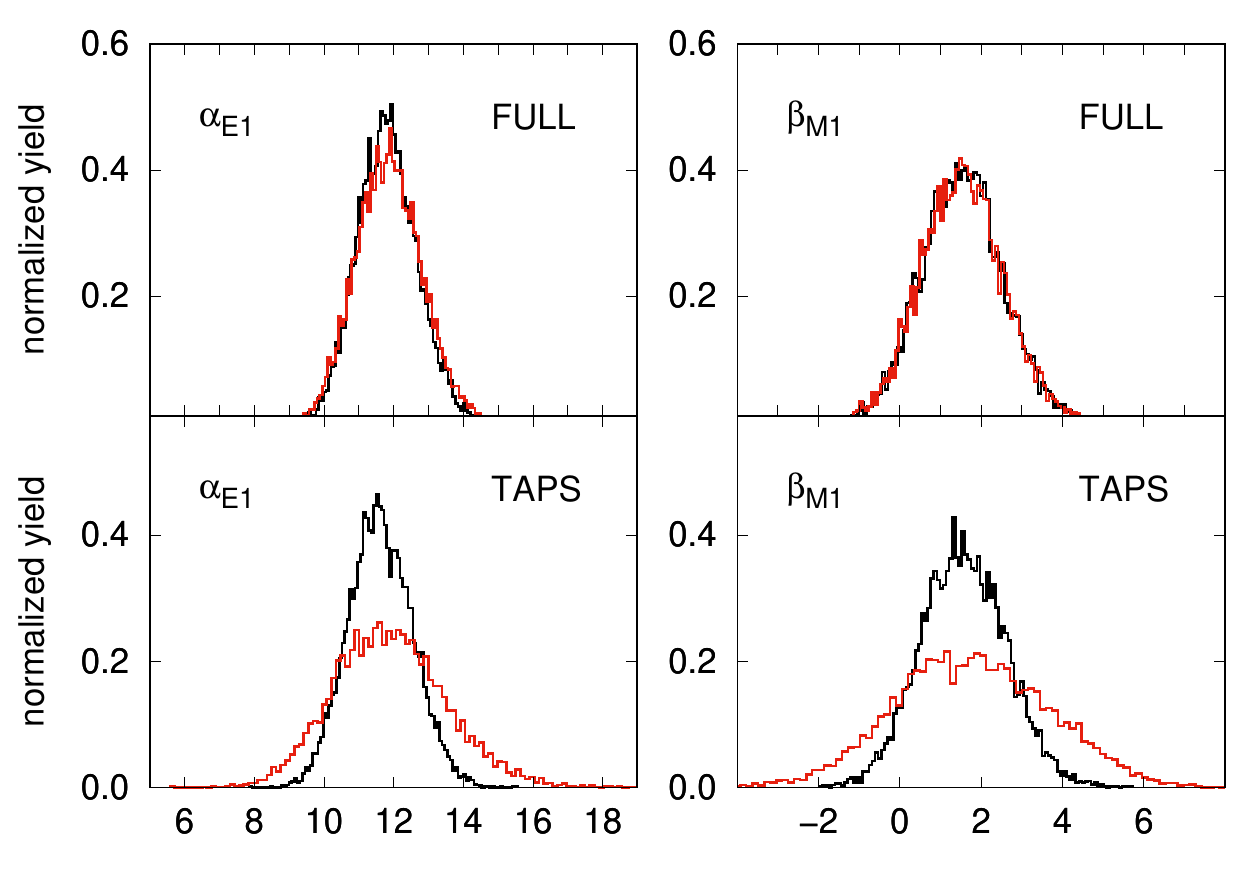}
\caption{Probability distributions of the fitted scalar dipole static polarizabilities $\astat$ (left panels) and $\bstat$ (right panels) in the Fit 2 (black curve) and Fit $2^\prime$ (red curve) conditions.
The results are obtained using the FULL data set (upper panels) and the TAPS data set (lower panels).\label{Fig2}}
	\end{figure}

\begin{figure}[h!]
\includegraphics[scale=1.]{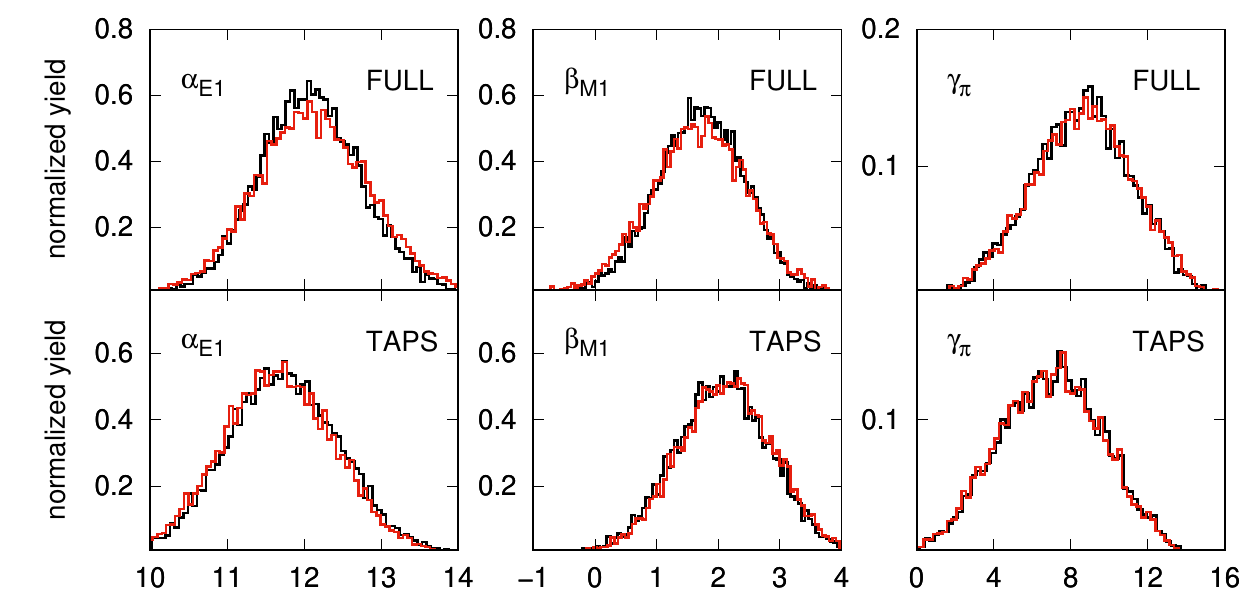}
\caption{Probability distributions of the fitted static polarizabilities $\astat$ (left panels) and $\bstat$ (central panels) and the backward spin polarizability $\gamma_\pi$ (right panels) in the Fit 3 (black curve) and Fit $3^\prime$ (red curve) conditions.
The results are obtained using the FULL data set (upper panels) and the TAPS data set (lower panels).\label{Fig3}}
	\end{figure}

We discuss in this section the results of the fit, performed under several configurations: 
	\beitem
\item Fit 1: with Baldin's sum rule, and systematic errors excluded: $\astat -\bstat$ as free parameter;
\item Fit $1^\prime$: with Baldin's sum rule, and systematic errors included: $\astat -\bstat$ as free parameter;
\item Fit 2: without Baldin's sum rule, and systematic errors excluded: $\astat$ and $\bstat$ as free parameters;
\item Fit $2^\prime$: without Baldin's sum rule, and systematic errors included: $\astat$ and $\bstat$ as free parameters;
\item Fit $3$: with Baldin's sum rule, and systematic errors excluded: $\astat -\bstat$ and $\gamma_\pi$ as free parameters;
\item Fit $3^\prime$: with Baldin's sum rule, and systematic errors included: $\astat -\bstat$ and $\gamma_\pi$ as free parameters.
	\enitem
All these different fits are performed using both the FULL and TAPS data sets.
The corresponding results
are summarized in \tref{tab:results} and shown in Figs.~\ref{Fig1}-\ref{Fig3}.
In all the cases, the probability distributions of the fit parameters are very similar to Gaussian functions.

 A few comments are in order:
\beitem
\item 
  the values of the fitted $\astat$ and $\bstat$ depend on the choice of the data set, but are all consistent within 
the uncertainties;
\item
  the sum of 
  the values of $\astat$ and $\bstat$ from the Baldin-unconstrained fit
  is well compatible, within the fit errors, with the Baldin's sum rule value 
\item the inclusion of systematic errors does not change the central values of the fitted parameters, but increases their uncertainties.
This effect is mostly visible for the TAPS data set fitted in the Fit 2 and Fit $2^\prime$ conditions, while it is reduced for the FULL data set, where
the effects of the systematic errors in the different subsets are, at least partially, compensated (see Figs.~\ref{Fig1}-\ref{Fig3}).
\item
  when the systematic errors are taken into account, the central values of the
  $\hat\chi^2/dof$ do not change. 
  However, the corresponding p-values significantly change for the FULL data
  set, since higher values
  of $\hat\chi^2/dof$ are more likely to occur.
  This effect is clearly visible from the cumulative distribution functions (CDFs) of $\hat \chi^2$ shown in
  \frtwo{fig:chi2cum}{fig:chi2cumnoerr}.
When we fit a single data set, as in the case of the TAPS data set, 
the systematic error becomes
 a common scale factor for all the data points and it
 does not change the p-value. Therefore, the main effect of the systematic-error propagation is to increase the statistical errors on the fitted parameters (see Fig.~\ref{Fig3});
\item the fitted values of $\gamma_\pi$ in the Fit $3^\prime$ conditions and with the additional contribution from the $\pi^0$-pole, i.e. $\gamma_{\pi}^{\text{tot}}=\gamma_\pi+\gamma_\pi^{\pi^0-\text{pole}},$ are in very good agreement with the values extracted within the fixed-$t$ unsubtracted DR analysis~\cite{Schumacher:2005an,Wolf:2001ha,Camen:2001st,OlmosdeLeon:2001zn}:
	\begin{align}
\text{LARA~\cite{Wolf:2001ha}}&: \quad {\gamma_\pi^{\text{tot}}} = -40.9 \pm 0.4\pm 2.2, & \text{SENECA~\cite{Camen:2001st}}&: \quad {\gamma_\pi^{\text{tot}}} = -39.1 \pm 1.2\pm 0.8 \pm 1.5, \nn
\text{TAPS~\cite{OlmosdeLeon:2001zn}}&: \quad {\gamma_\pi^{\text{tot}}} = -35.9 \pm 2.3, & {\text{Fit 3}}^\prime (\text{FULL})&: \quad {\gamma_\pi^{\text{tot}}} = -38.11^{+2.85}_{-2.94}.
	\end{align}
The results of Refs.~\cite{Wolf:2001ha,Camen:2001st} are obtained using data above the pion-production threshold, while  the result of Ref.~\cite{OlmosdeLeon:2001zn} is extracted from the complete TAPS data set, ranging up to photon energies of $165$ MeV.
	\enitem
	\betab[t!]
	\renewcommand{\arraystretch}{1.7}
\begin{tabular}{|c|c|c|c|c|}
\hline
\multicolumn{5}{|c|}{FULL data set}\\
\hline
\hline
fit conditions & $\astat$ & $\bstat$ & $\gamma_\pi$ & $\hat\chi^2/dof$ (p-value)\\
\hline
Fit1 & $12.00 ^{+0.41}_{-0.47}$ & $1.80^{+0.46}_{-0.48}$ & fixed & $1.25$ ($3\%$)\\
Fit $1^\prime$ & $12.03^{+0.48}_{-0.54}$ & $1.77^{+0.52}_{-0.54}$ & fixed & $1.25$ ($12\%$)\\
Fit 2 & $11.82^{+0.81}_{-0.91}$ & $1.54^{+0.95}_{-1.00}$ & fixed & $1.26$ ($4\%$)\\
Fit $2^\prime$ & $11.86^{+0.93}_{-0.99}$ & $1.54^{+0.95}_{-1.05}$ &fixed & $1.26$ ($13\%$)\\
Fit 3 & $12.08^{+0.61}_{-0.64}$ & $1.71^{+0.70}_{-0.75}$ & $8.52^{+2.72}_{-2.93}$ & $1.26$ ($4\%$)\\
Fit $3^\prime$ & $12.12^{+0.68}_{-0.77}$ & $1.68^{+0.77}_{-0.79}$ & $8.59^{+2.85}_{-2.94}$ & $1.26$ ($13\%$) \\
\hline
\hline
\multicolumn{5}{|c|}{TAPS data set}\\
\hline
\hline
fit conditions & $\astat$ & $\bstat$ & $\gamma_\pi$ & $\hat\chi^2/dof$ (p-value)\\
\hline
Fit 1 & $11.87^{+0.50}_{-0.54}$ & $1.93^{+0.52}_{-0.56}$ & fixed &$1.32$ ($7\%$) \\
Fit $1^\prime$ & $11.82^{+0.50}_{-0.55}$ & $1.98^{+0.52}_{-0.56}$ & fixed & $1.32$ ($7\%$)\\
Fit 2 & $11.62^{+0.86}_{-0.96}$ & $1.57^{+1.01}_{-1.04}$ & fixed & $1.34$ ($8\%$)\\
Fit $2^\prime$ & $11.89^{+1.50}_{-1.67}$ & $1.76^{+1.83}_{-2.00}$ & fixed &$1.34$ ($8\%$)\\
Fit 3 & $11.74^{+0.68}_{-0.78}$ & $2.06^{+0.77}_{-0.84}$ & $6.95^{+2.81}_{-3.09}$ & $1.34$ ($9\%$)\\
Fit $3^\prime$ & $11.67^{+0.68}_{-0.77}$ & $2.12^{+0.76}_{-0.78}$ & $6.85^{+2.83}_{-3.07}$ & $1.34$ ($9\%$)\\
\hline
\end{tabular}
\caption{Results of the fits for the static polarizabilities $\astat$, $\bstat$ and $\gamma_\pi$ using the FULL and TAPS data sets and different fit conditions, together with the corresponding $\hat\chi^2/dof$ and p-values.
}\label{tab:results}
	\entab
	
	\begin{figure}[t!]
  \begin{subfigure}[b]{0.45\textwidth}
    \includegraphics[scale=1.]{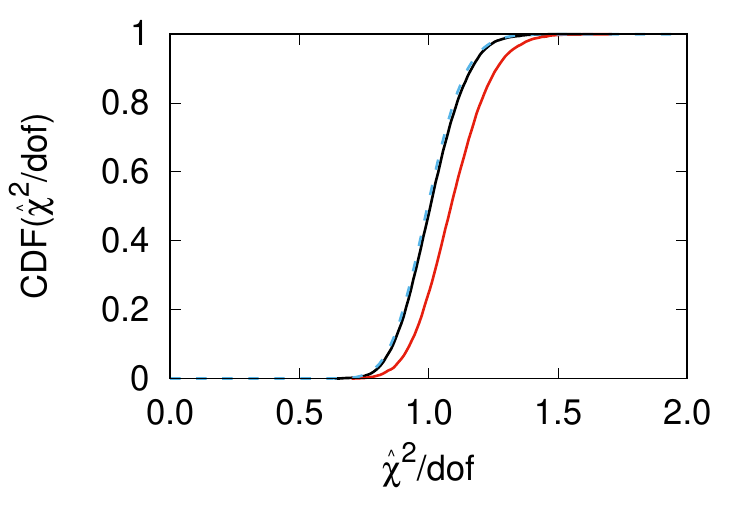}
    \subcaption{ \hspace{-0.8 truecm} }
    \label{fig:chi2cum_Ffit1}
  \end{subfigure}
  \centering
  \begin{subfigure}[b]{0.45\textwidth}
  \centering
    \includegraphics[scale=1.]{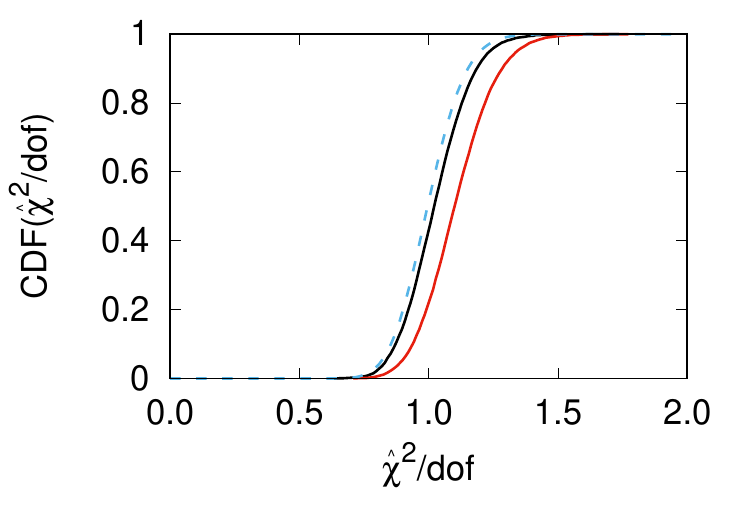}
  \centering
  \subcaption{ \hspace{-1 truecm} 
    \label{fig:chi2cum_Ffit3}}
  \end{subfigure}
  \caption{Cumulative distribution functions for the variable $\hat \chi^2/dof$ in the case of Fit 1 (left panel, black curve), Fit $1^\prime$ (left panel, red curve), Fit 3 (right panel, black 
  curve), Fit $3^\prime$ (right panel, red curve), using the FULL data set. The dashed-blue curves are the cumulative distribution functions of a pure reduced $\chi^2$. \label{fig:chi2cum}}
  \end{figure}

\begin{figure}[h!]
  \centering
  \begin{subfigure}[b]{0.45\textwidth}
    \includegraphics[scale=1.]{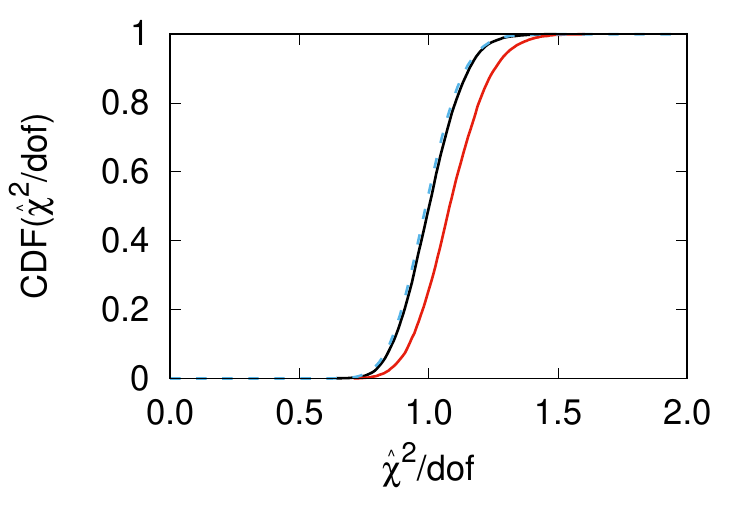}
    \caption{}
    \label{fig:chi2cum_Ffit1}
  \end{subfigure}
  \begin{subfigure}[b]{0.45\textwidth}
    \includegraphics[scale=1.]{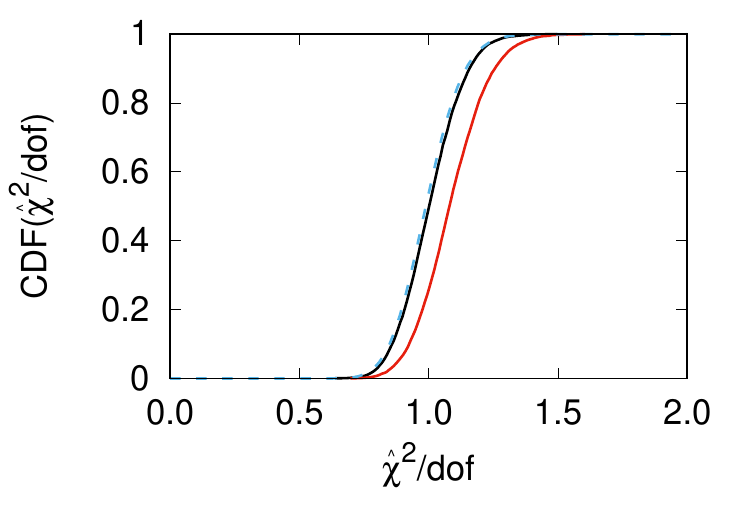}
    \caption{}
    \label{fig:chi2cum_Ffit3}
  \end{subfigure}
  \caption{The same as in \fr{fig:chi2cum}
    but neglecting the errors 
    on the polarizability values not treated as free parameters in the fit procedure.
    \label{fig:chi2cumnoerr}} 
  \end{figure}

From all the  results we conclude        
that the inclusion of the systematic errors in the fitting procedure is very important
since,  in this case,
the p-value associated to the $\hat \chi^2$ changes significantly (see \fr{fig:chi2cum}),
while the uncertainty on the fitted parameters changes in a much less pronounced way.
This behavior can be observed only thanks to the bootstrap method, since it is not possible to compute the correct
p-values without resorting to the Monte Carlo replicas, once included the systematic errors.

As mentioned before, the $\hat\chi^2$ parameter
obtained in the different fitting conditions is never distributed like a  $\chi^2$ probability function.
  This effect is due to the correlation present among all the points in each subset, and is clearly visible from the CDFs shown in \fr{fig:chi2cum}.
\section{Consistency checks on the available data set}
\label{sec:discussion}

The scientific community has not reach so far a common agreement on the definition of the data set of proton RCS below pion-production threshold~\cite{Pasquini:2017ehj,Griesshammer:2012we,Krupina:2017pgr}.
As pointed out in Ref.~\cite{Krupina:2017pgr} and in \secref{subsect:results} of this work, the values obtained from
a fit of $\astat$ and $\bstat$ strongly depend on the choice of the data set.
In this section, we apply a few basic statistical tests to investigate the consistency of the data set and the possible occurrence of outliers. We will discuss the FULL data set, the TAPS data set and the SELECTED data set, as an example of selection from the complete data set, all below the pion-production threshold as detailed in App.~\ref{sec:data}.

In this section, we will  use the standard minimization technique and we will not take systematic errors into account,
in order to work in a well-established fitting condition and investigate the pure statistical features of the experimental data. We set the Fit 1 condition (\ie we use $\astat-\bstat$ as free parameter and we neglect the systematic errors) and we use the conventional $\chi^2$ of \eqr{eq:chiboot3} as minimization function, without implementing the bootstrap procedure. 
Therefore, the errors on the fixed spin polarizabilities are not included in the results of the tests, while the uncertainties of $\astat + \bstat$ affect the electric and magnetic polarizabilities errors as $\epsilon_{\astat,\bstat} = \sqrt{(\epsilon_{\astat+\bstat}^2 + \epsilon_{\astat-\bstat}^2 )}/2$. We will refer to this fitting configuration as test-fit. The result of the test-fit applied to the FULL data set leads to the best values $\astat = 11.99 \pm 0.31$ and $\bstat = 1.81\pm 0.31$, which are almost identical to the values given in \tref{tab:results}, with the Fit 1 condition applied to the FULL data set. The
tiny difference in the central values and the different statistical errors are due to the propagation of the uncertainties of those polarizabilities that are not treated as free parameters in the bootstrap fit.
	\subsection{The Jackknife resampling}
	\label{sec:jack}
        A possible strategy to discuss the consistency of the data set is the Jackknife, a resampling technique that 
        can be considered as 
        a particular case of the non-parametric bootstrap technique.
        Given a data set $D=\{d_i\}, i=1,\dots,n$, composed by $n$ points, we can define $n$ data subsets by removing one datum
        at a time, \ie $D_k = D\setminus \{d_k\}$, where $k=1,\dots,n$.
We then fit the model $T(p)$ to every $D_k$ data set, obtaining a best value of the parameters $\hat p_k$ for each set. From the $n$-tuple of $\hat p_k$, we can compute the average $\hat p_{Jack}$ and its sample standard deviation $\sigma_{Jack}$. 
An outlier $k$ 
is expected to give a result far from the average value, \ie $|\f{\hat{p}_{k}-\hat p_{Jack}}{\sigma_{Jack}}|\gg 1$. Instead, if there are no evident outliers, we expect that all the variables $\hat p_k$ follow, at least approximately,
Gaussian confidence levels~\cite{ref:giacomo}.
In this way, we can identify possible deviations of a data subset from the other ones.

\begin{figure}
  \centering
  \begin{subfigure}[b]{0.45\textwidth}
    \includegraphics[scale=1.]{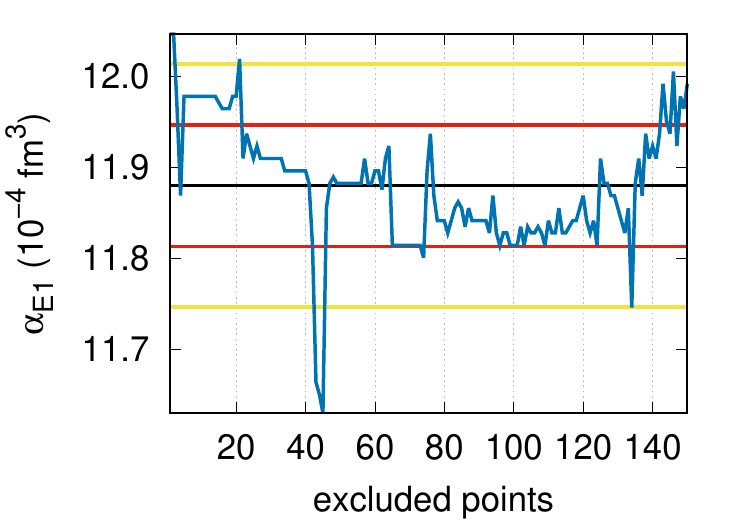}
    \label{fig:jack_alpha_full}
  \end{subfigure}
  ~ 
  \begin{subfigure}[b]{0.45\textwidth}
    \includegraphics[scale=1.]{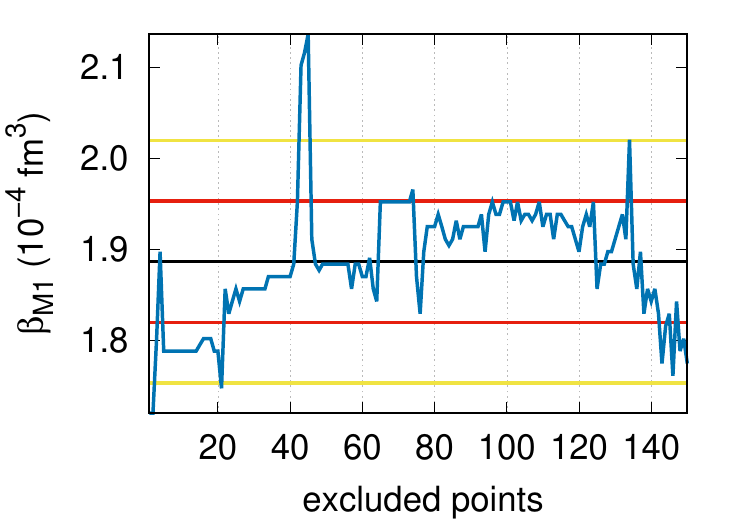}
    \label{fig:jack_beta_full}
  \end{subfigure}\\
    \begin{subfigure}[b]{0.45\textwidth}
    \includegraphics[scale=1.]{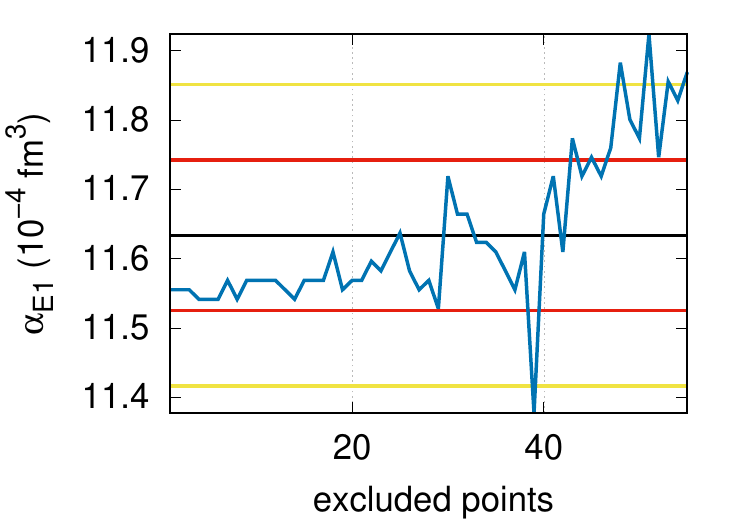}
    \label{fig:jack_alpha_full}
  \end{subfigure}
  ~ 
  \begin{subfigure}[b]{0.45\textwidth}
    \includegraphics[scale=1.]{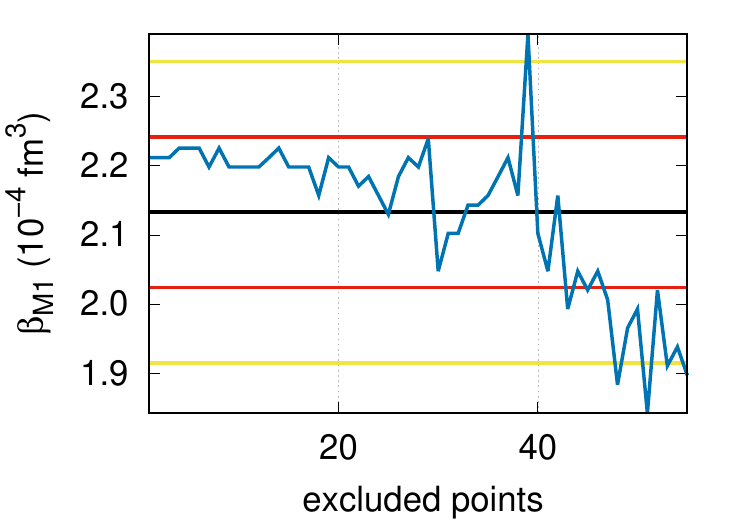}
    \label{fig:jack_beta_full}
  \end{subfigure}\\
    \begin{subfigure}[b]{0.45\textwidth}
    \includegraphics[scale=1.]{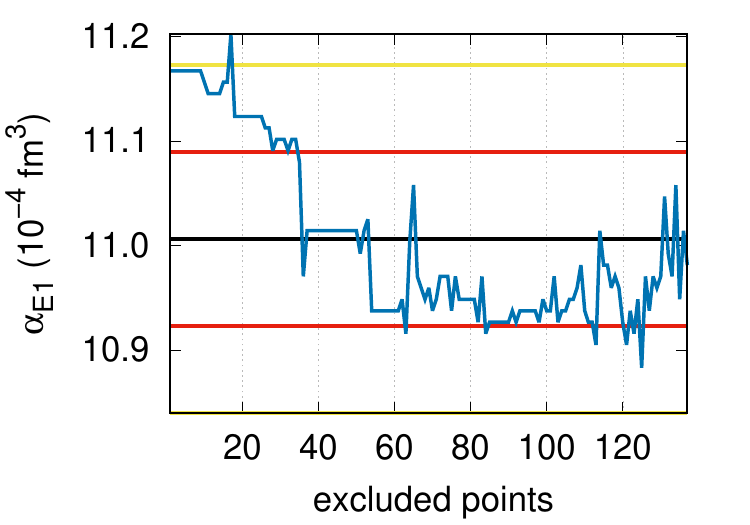}
    \label{fig:jack_alpha_full}
  \end{subfigure}
  ~ 
  \begin{subfigure}[b]{0.45\textwidth}
    \includegraphics[scale=1.]{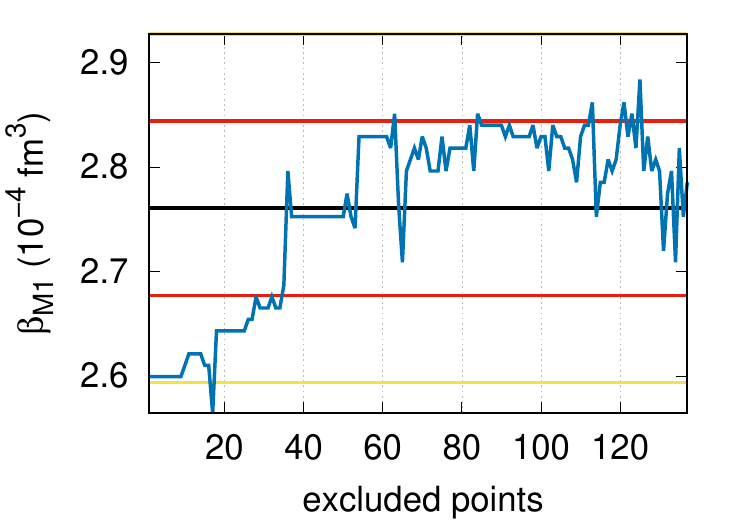}
    \label{fig:jack_beta_full}
  \end{subfigure}  
  \caption{Results from the Jackknife (blue line) for $\alpha_{E1}$ (left panels) and $\beta_{M1}$ (right panels). The red (yellow) lines correspond to
  the 1 $\sigma_p$ (2 $\sigma_p$) sample standard deviations. From top to bottom: results for the FULL data set, TAPS data set and the SELECTED data set.}\label{fig:jack}
\end{figure}

We apply the Jackknife to the FULL, TAPS and SELECTED data sets: the best values of $\astat$ and $\bstat$ versus the index $k$ of the excluded point in each subset are plotted in Fig.~\ref{fig:jack}. In the case of the FULL data set, we note that the statistical fluctuations are well in agreement with the expected Gaussian confidence levels ($\sim95\%$ of the occurrences within the $2\sigma$ range). We can then conclude that there is no clear evidence of outliers.

In the case of the SELECTED data set, we obtain very similar results,
with less pronounced fluctuations ($\sim98\%$ of the occurrences within the $2\sigma$ range).
This does not necessarily implies that there is an improvement in the data set. Instead,
this behavior may simply reflect the fact that the data points excluded from the set are
not "close enough to'' the model predictions.

The same test applied to the TAPS data set shows a clear dependence of the values of $\astat$ and $\bstat$ on the scattering angle. This feature is due to the fact that
the data are ordered by increasing scattering angles and that the sensitivity of the unpolarized RCS cross section to $\astat-\bstat$ is higher in the backward scattering region: when a single datum is removed in the backward region, the value of $\bstat$ decreases and $\astat$ increases.

	\subsection{Residual analysis}
			\label{sec:residual}
In order to cross-check the stability of the FULL data set, we performed the analysis of the residuals, defined as
	\beq
	\label{eq:residuals}
\xi_i \equiv \f{E_i - \hat T_i}{\sigma_i},
	\enq
        where $E_i$ is the $i^{th}$ experimental datum with the uncertainty $\sigma_i$, and $\hat T_i$ is the model prediction obtained with the best values of the fitted parameters. If the model is able to correctly describe the experimental datum, the value $E_i$ can be considered a possible outcome of the probability distribution of $T_i$. In this case, the variable $\xi_i$ of \eqr{eq:residuals} is Gaussian distributed as $\mc G[0,1]$.

The residual analyses for the FULL and SELECTED data sets are shown
in \fr{fig:residuals}, together with the q-q plots, representing the CDF($\xi_i$) vs CDF($z$), with $z$ a Gaussian distributed variable according to $\mc N[0,1]$.
The variable $\xi_i$
has mean value and standard deviation in
fairly good agreement with the expectations.
In the case of the SELECTED data set, we observe again less pronounced statistical fluctuations mostly due to the  exclusions of the subsets
1~\cite{Oxley:1958zz}  and 7~\cite{Baranov:1975ju}.
This is also shown by the fact that  the CDF($\xi_i$) for the SELECTED data set approaches the maximum value of unity faster  than in the case of the FULL  data set.

More precisely, the FULL data set shows three points lying outside
  the $3\sigma$ range: this configuration can happen with a $1\%$ probability, assuming only Gaussian random choice.
  Such a value, even if not extremely low, points to possible outliers inside
  the data set.
  On the other hand, the SELECTED data set has only 2 points outside the $2\sigma$ range; in this case the associated probability is $\simeq 3 \%$.
Such a low value seems to indicate that at least
  some of the discarded points were not outliers.

 The FULL and SELECTED data sets  have a very similar statistical significance  
 and this ambiguity  
  can be only resolved with new sets of
  precise and accurate data, especially  in the in the backward  angular region where the sensitivity to $\astat-\bstat$ is higher.

    All these conclusions are in agreement with the results obtained with the bootstrap method and shown in \tref{tab:results}. Under the Fit 1 condition, taking only into account the statistical (Gaussian) errors, the p-value of the fit is about $3\%$, which is very close to the $1\%$ occurrence probability.
    However, when the systematic errors are included in the fitting procedure (Fit 1$^\prime$), the statistical significance strongly increases ($12 \%$), thus indicating that the occurrence probability of the FULL data set is higher,
   when taking properly into account all the data error sources.
   Moreover, since there is not a clearly identified source of possible experimental problems that could affect the data discarded  in the SELECTED data base, we prefer not to exclude any point
   to keep the highest sensitivity to $\astat-\bstat$. Instead, we propose   to treat the suspicious points with the approach outlined below.

\begin{figure}
  \centering
  \begin{subfigure}[b]{0.45\textwidth}
    \includegraphics[scale=.8]{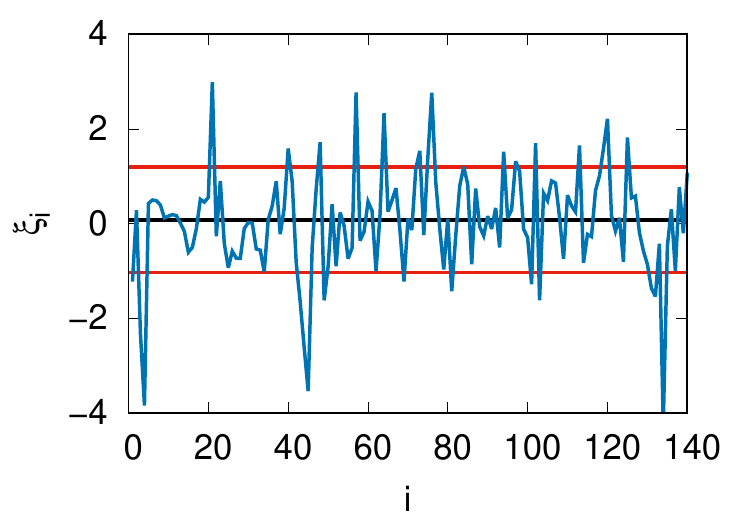}
    \caption{}
    \label{fig:res_Ffit1}
  \end{subfigure}
  ~ 
  \begin{subfigure}[b]{0.45\textwidth}
 \includegraphics[scale=.8]{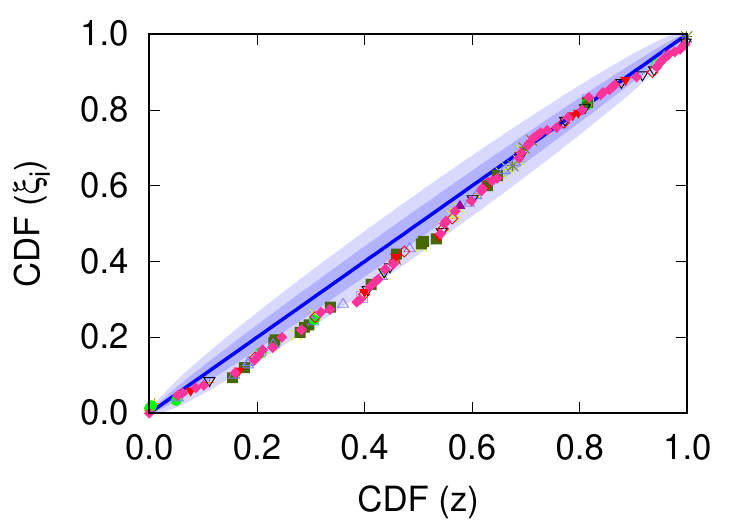}
    \caption{}
    \label{fig:lag_Ffit1}
  \end{subfigure}\\
    \begin{subfigure}[b]{0.45\textwidth}
\includegraphics[scale=.8]{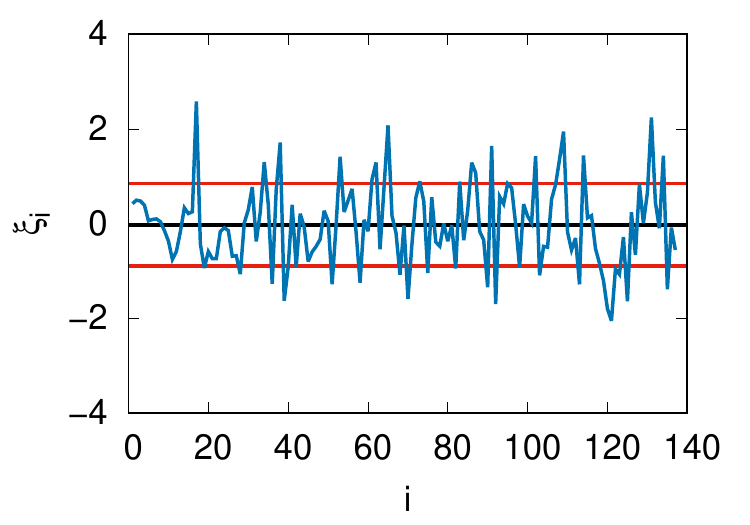}
    \caption{}
    \label{fig:res_gries}
  \end{subfigure}
  ~ 
  \begin{subfigure}[b]{0.45\textwidth}
 \includegraphics[scale=.8]{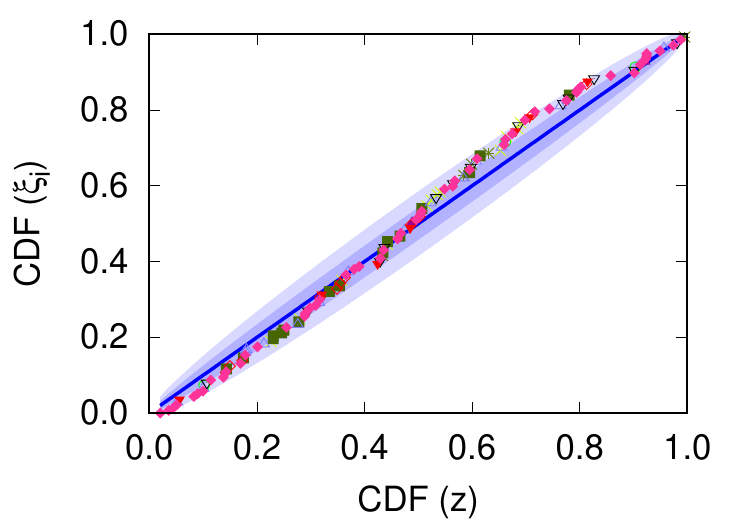}
    \caption{}
    \label{fig:lag_gries}
  \end{subfigure}
  \caption{Residual analysis applied to the FULL (top panels) and SELECTED (lower panels) data set . The left panels show the values of $\xi_i$ (blue curves), with their mean value (black curves) and their sample standard deviation band (red curves).
    The right panels are the q-q plots of $\xi_i$ compared with the results 
    expected in the case of a normal distribution (diagonal blue line).
    The dark (light) blue band shows the $1\sigma$ (2$\sigma$) uncertainty region due to the
      data set dimension.
    The labels of the data sets are described in \apref{sec:data}.}\label{fig:residuals}
\end{figure}
	\subsection{The $\chi^2$ per set}
		\label{sec:chi2set}

Given a data set composed by subsets with $n_{set}$ points, we can define for each subset the following variable
	\beql{eq:chi2set}
\chi^2_{set} \equiv \f{1}{n_{set}}\sum_{i=1}^{n_{set}} \left(\f{E_i - \hat T_i}{\sigma_i}\right)^2.
\enq

If the model $\hat T_i$ is able to well describe the data,
all the $\chi^2_{set}$ values should be fairly close to one. 
Viceversa, if $\chi^2_{set}\gg 1$, we cannot automatically
deduce that a data subset should be excluded. 
This parameter is evaluated using the particular model used in the fit procedure
  and a bias may be introduced
  by using large values of $\chi^2_{set}$ as criterion to exclude data sets.

We applied this kind of analysis to the FULL data set and the results are shown in \fr{fig:chi2contrib}.
We can notice that most of the subsets
have $\chi^2_{set}\approx 1$,
while the subsets 1~\cite{Oxley:1958zz}
and 7~\cite{Baranov:1975ju} give
higher $\chi^2_{set}$ values.
As mentioned before,
these subsets are indeed excluded in the definition of the SELECTED data set.
However, both data sets have only 4 points each and with
such a small number of points
we can not exclude the occurrence of  pure statistical fluctuations,
  as mentioned in the previous section.

 An alternative method, first suggested in~\cite{ref:birge},
 is to rescale the statistical errors of the points of each data subset by a factor $\sqrt{\chi^2_{set}}$ and to repeat again
  the fit procedure (see also~\cite{ref:statbook}).
    This relies on  the assumption that a large $\chi^2_{set}$ value indicates underestimated measurement
    uncertainties that should be equally attributed to all the points of a given subset.    
  We then obtain new values for the fitted parameters ${\hat p}^\prime$ with the minimum of the $\chi^2$ function equal to 1,
  by construction.
  This strategy is again model dependent, but it can be used as an indication
  for the identification of outliers.
If there are no data subsets that behave as outliers and then could determine very different values for the fitted parameters, we would expect that ${\hat p}^\prime \simeq \hat p$. 

In our case,
the values of the fitted parameters obtained from the FULL data set with and without rescaling of the statistical errors are consistent within
the (large)  fit errors, i.e.
	\begin{align}
\text{no rescaling}:& \qquad\astat -\bstat = 10.17 \pm 0.47,\label{eq:no-rescaling}\\
\sqrt{\chi^2_{set}} \text{ rescaling}:& \qquad \astat -\bstat = 9.36 \pm 0.50.\label{eq:rescaling}
	\end{align}
        If we exclude from the fit the subsets 1 and 7,  without rescaling the errors, we obtain the value $\astat -\bstat = 9.01 \pm 0.50$, which is 
         very similar to the result in Eq.~\eqref{eq:rescaling} obtained with the rescaling method.
         As a matter of fact, the rescaling method is equivalent to reduce the impact of the data points in the subsets 1 and 7, since these points are weighted by their relatively high $\sqrt{\chi^2_{set}}$ factor. However, the rescaling does not lead to a significant improvement in the accuracy of the extraction, since     
         the error bars  in Eqs.~\eqref{eq:no-rescaling}  and \eqref{eq:rescaling} are very similar.         
         Furthermore, the difference between the central values in  Eqs.~\eqref{eq:no-rescaling} and \eqref{eq:rescaling} can be related to the angular distribution of the experimental data of sets 1 and 7, which is mainly in the backward scattering region, where the sensitivity to $\astat-\bstat$ is higher. 
         
	Given all these findings, we once more conclude that
        there is no clear evidence that these sets are outliers that should be
        excluded from the fit.

		\begin{figure}[]
\includegraphics[scale=.7]{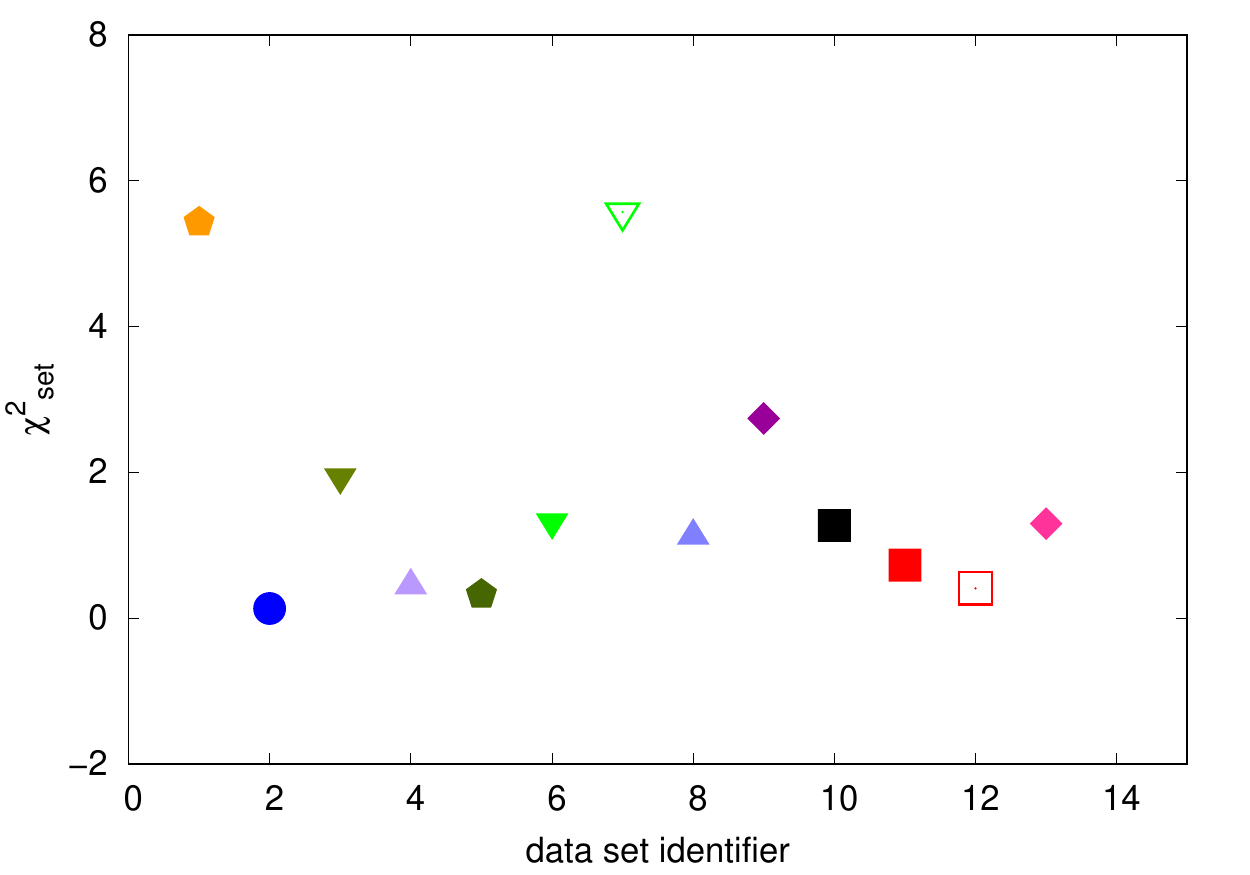}
\caption{$\chi^2_{set}$ term for each data sub set of the FULL set. The labels of the data sets are described in \apref{sec:data}.}\label{fig:chi2contrib}
	\end{figure}
%
\subsection{Behavior of the minimization function}
\label{sec:chi2profile}
In order to investigate the effect on the fit results of the exclusion of some data points,
we examined the  behavior of the minimization function versus
  the values of the fit parameters $\astat$ and $\bstat$.

The results for the FULL data set (with and without rescaling the statistical error by a factor $\sqrt{\chi^2_{set}}$), for the TAPS data set and for the SELECTED data set~\cite{Griesshammer:2012we} are shown in \fr{fig:chi2prof}.

When outliers are discarded from the fit, 
we would expect 
a significant reduction of the minimum of the $\chi^2$ function as well as  a more pronounced convexity, corresponding to
 smaller errors for the fitted parameters. 
In  the case of the SELECTED data set, we indeed observe that the minimum value of the reduced $\chi^2$ function
is closer to 1,
but the shape of the  minimization function is the same as in the case of the FULL data set,
\ie the errors on the fitted parameters remain ultimately the same.
 
	\begin{figure}[t]
	\centering
	   \begin{subfigure}[b]{0.45\textwidth}
 \includegraphics[scale=1.4]{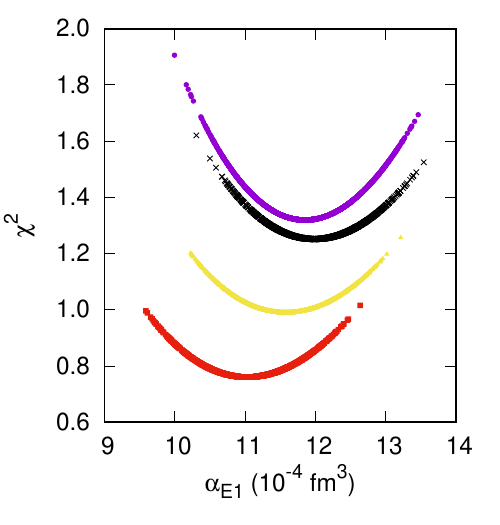}
    \caption{}
    \label{fig:res_gries}
  \end{subfigure}
  \begin{subfigure}[b]{0.45\textwidth}
 \includegraphics[scale=1.4]{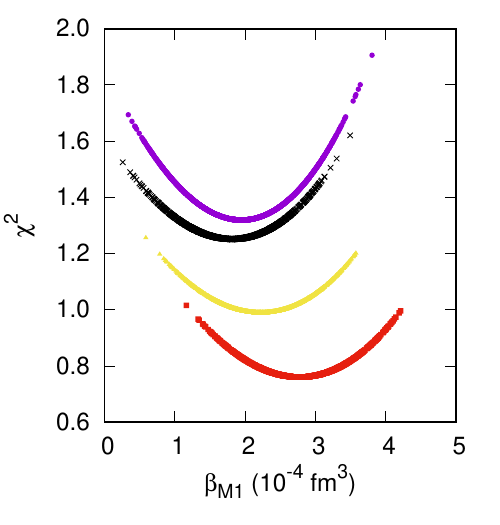}
    \caption{}
    \label{fig:lag_gries}
  \end{subfigure}
\caption{The $\chi^2$ profile as function of $\astat$ (a) and $\bstat$ (b).
The black curves are the results for the original FULL data set, while the yellow curves correspond to the results for the FULL data set with the $\sqrt{\chi^2_{set}}$ rescaling of the statistical errors.
The purple and red curves show, respectively, the results for the TAPS  and the SELECTED data set.}\label{fig:chi2prof}
 \end{figure}
        
This simple analysis gives another additional hint
        that there is no clear evidence of the presence of outliers 
and no strong enough motivation for the exclusion of some data points from the FULL data set.


        \subsection{Summary of the tests}

All the previous consistency tests led us to the conclusion that there are no strong motivations for the exclusion of any data point
from the global RCS data set below pion-production threshold, even though
we observed significant deviations
for a few data points at the backward scattering angles.
The residual analysis lets us to conclude that the FULL and the SELECTED data sets have almost the same statistical significance and also the fit errors are basically the same in both cases.

  An alternative approach to handle the suspicious points is to rescale their statistical errors by a factor $\sqrt{\chi^2_{set}}$ rather than exclude them from the data set. 
  Also the bootstrap fitting technique is useful in these cases to check if the inclusion of  systematic errors in the fitting procedure improves the  
  significance of the obtained results.

Since most of these points are located in the  backward scattering region,
  where the RCS unpolarized cross section has the larger sensitivity to $\astat-\bstat$, their exclusion may lead to biased results.
  
We conclude that the main reason of the sizeable uncertainties 
    that are present at the moment in the extraction of the
  scalar polarizabilities and especially of $\beta_{M1}$ 
are mainly due to  the intrinsic limitations (poor accuracy and
 scarcity) of the data set at our disposal.

\section{Available extractions of RCS scalar dipole static polarizabilities}
\label{sec:summary}

In \fr{fig:summary}, we collect the available results for the extraction of the scalar dipole static polarizabilities from RCS at low energies.
The red solid curve show the results from this work, obtained from the bootstrap-based fit using the FULL data set with the constraint of the Baldin's sum rule
and taking into account the effects of the systematic errors of the experimental data and the propagation of the statistical errors of the fixed polarizabilities $\astat+\bstat$, $\gamma_0$, $\gamma_\pi$, $\gamma_{E1E1}$ and $\gamma_{M1M1}$ (Fit $1^\prime$ conditions). Within our fitting technique, we are able to evaluate
the correlation coefficient $ \rho_{\astat-\bstat}$ among $\astat$ and $\bstat$: this determines the ellipse-shape in \fr{fig:summary}. All the other correlation terms are given in \apref{sec:correlations}.
Numerically, we obtain the following best values with the 68\% confidence-level error bar
\beql{eq:summary}
 \astat =12.03^{+0.48}_{-0.53},\qquad \bstat =1.77^{+0.52}_{-0.54}, \qquad\hat\chi^2 =1.25 \quad (\text{p-value}=12\%),\quad \rho_{\astat-\bstat}=-0.72.
	\enq
These results are in very good agreement with the ones obtained using a traditional $\chi^2$ fitting procedure in a fixed-$t$ subtracted DRs framework~\cite{Drechsel:2002ar}.
The experimental fits shown by black curves have been obtained within unsubtracted DRs~\cite{MacGibbon:1995in,Federspiel:1991yd,OlmosdeLeon:2001zn}.
The light-green band shows the experimental constraint 
on the difference $\astat-\bstat$ from Zieger et al.~\cite{Zieger:1992jq}.
The green solid curve shows the B$\chi$PT predictions of Ref.~\cite{Lensky:2015awa}.
The blue solid curve corresponds to the $68\%$ ellipse of the Baldin constrained fit of Ref.~\cite{McGovern:2012ew,Griesshammer:2015ahu}, using the SELECTED data set and the HB$\chi$PT framework. These results are in excellent agreement also with the fit within B$\chi$PT of Ref.~\cite{Lensky:2014efa}. 
We also show the latest value from PDG~\cite{Patrignani:2016xqp} (solid black disk):
\begin{eqnarray}
\alpha_{E1} = 11.2 \pm 0.4,\qquad
\beta_{M1} = 2.5 \pm   0.4.
\end{eqnarray}
They differ from the 2012 and earlier editions by inclusion of the data fit analysis within HB$\chi$PT~\cite{McGovern:2012ew}.

We note that there is a discrepancy
between the values obtained in the framework of effective field theories~\cite{McGovern:2012ew,Griesshammer:2012we,Lensky:2015awa}
and the results obtained using DRs, even if they are compatible within the $2\sigma$-range.
		\begin{figure}[t]
	\centering
	 \savebox{\imagebox}{ \includegraphics[scale=1.2]{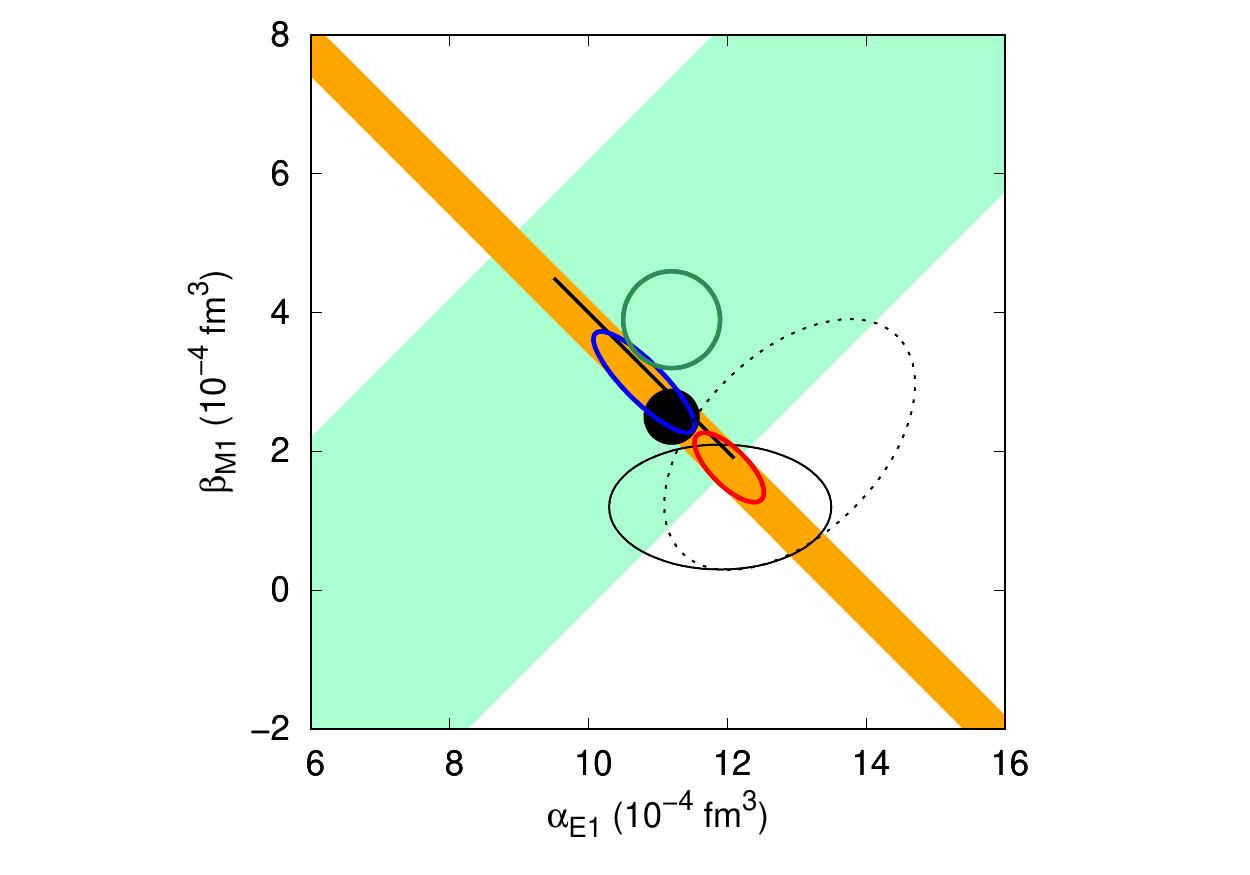}}%
	   \begin{subfigure}[t]{0.65\textwidth}
	     \centering\usebox{\imagebox}
  \end{subfigure}
  \begin{subfigure}[t]{0.25\textwidth}
    \centering\raisebox{\dimexpr\ht\imagebox-\height}{
 \includegraphics[scale=0.8]{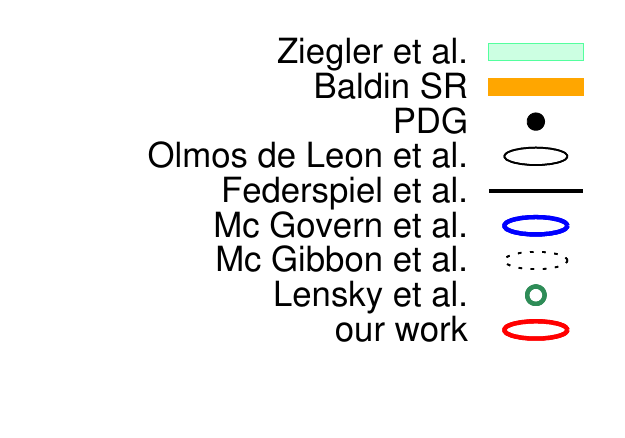}}
  \end{subfigure}
\caption{Results for $\astat$ vs $\bstat$ obtained in different frameworks. The light-green band shows the experimental constraint 
on the difference $\astat-\bstat$ from Zieger et al.~\cite{Zieger:1992jq}, while the orange band is the average over the available Baldin's sum rule evaluations~\cite{Hagelstein:2015egb}.
The experimental extractions are from 
Federspiel et al.~\cite{Federspiel:1991yd} (straight black line), obtained from the fit of $\astat-\bstat$ constrained by $\astat+\bstat = 14.0$; MacGibbon et al.~\cite{MacGibbon:1995in} (short-dashed black curve, unconstrained fit); TAPS~\cite{OlmosdeLeon:2001zn} (solid black curve, unconstrained fit).
The green solid curve is the B$\chi$PT prediction from Ref.~\cite{Lensky:2015awa}, while the blue solid curve shows the fit with the constraint $\astat+\bstat = 13.8 \pm 0.4$ within HB$\chi$PT from Refs. ~\cite{McGovern:2012ew,Griesshammer:2015ahu}. 
The solid black circle shows the PDG results~\cite{Patrignani:2016xqp}.
The solid red curve is the extraction from this work (Fit 1$^\prime$), using fixed-$t$ subtracted DRs. 
 \label{fig:summary}}
	\end{figure}
In order to shed some light on the origin of 
the difference between the results from the extraction within HB$\chi$PT and fixed-$t$ subtracted DRs,
we performed some test-fits, in the condition described in \secref{sec:discussion}, using  fixed-$t$ subtracted DRs with input from the central values of HB$\chi$PT predictions for the spin polarizabilities.
The results for the leading-order spin polarizabilities in HB$\chi$PT read~\cite{McGovern:2012ew,Griesshammer:2015ahu} $\gamma_{E1E1}=-1.1 \pm 1.9$, $\gamma_{M1M1}=2.2 \pm0.5 \text{(stat)}\pm 0.6$, $\gamma_{0}=-2.6\pm0.5 \text{(stat)}\pm 1.8$, and $\gamma_{\pi}=5.6\pm0.5 \text{(stat) }\pm 1.8$, and are quite different from the experimental values used in our DR analysis. On top of that, we noticed
a different evaluation for the $\pi^0$-pole contribution calculated in Ref.~\cite{McGovern:2012ew}, which is $-45.9$ for $\gamma_\pi$.
In \tref{tab:summary}, we compare the test-fit values for $\astat$ and $\bstat$ in the case we use the results of the spin polarizabilities and the $\pi^0$pole 
from the experimental extraction~\cite{Martel:2014pba} or the corresponding values from HB$\chi$PT~\cite{McGovern:2012ew,Griesshammer:2015ahu}, with the $\pi^0$-pole contribution reported in~\cite{McGovern:2012ew} (results in brackets). This analysis has been performed for both the FULL and SELECTED data sets, in order to investigate the dependence of the results not only on the values of the spin polarizabilities, but also on the choice of the data set (see Ref.~\cite{Krupina:2017pgr} for a more comprehensive discussion). If we focus on the central values of $\bstat$, we notice that the different input for the spin polarizabilities affects the results by 20-30$\%$, while the choice of the data set leads to a 40-50$\%$ increase.
It is certainly too simplistic to estimate the model dependence of the two extractions with the different values of the spin polarizabilities.
However, in the energy range below pion production threshold, this gives a rather good indication of the main effects due to the model dependence.

		\betab[t]
	\renewcommand{\arraystretch}{1.7}
\begin{tabular}{|c|c|c|}
\hline
 & FULL & SELECTED \\
\hline 
$\astat$ & $11.99 \pm 0.31 \qquad ( 11.47\pm 0.30)$ & $11.02 \pm 0.33\qquad (10.46 \pm 0.32)$\\ 
$\bstat$ & $1.81 \pm 0.31 \qquad ( 2.33 \pm 0.30 )$ & $ 2.78 \pm 0.33 \qquad (3.34 \pm 0.32 )$\\ 
\hline
\end{tabular}
\caption{Results for $\astat$ and $\bstat$ from the test-fit of the FULL and the SELECTED data set, and taking different values for the leading-order spin polarizabilities:
the experimental results from Ref.~\cite{Martel:2014pba} and the values predicted in HB$\chi$PT~\cite{Griesshammer:2012we} (results in brackets).}\label{tab:summary}
	\entab
The results for the RCS differential cross section obtained with the values of \eqr{eq:summary} for the scalar dipole polarizabilities and the 
experimental values of Ref.~\cite{Martel:2014pba} for the leading-order spin polarizabilities 
are shown in \fr{fig:cross} as a function of the lab photon energy $E_\gamma$ and the lab scattering angle  $\theta_{\text{lab}}$, in comparison with the experimental data of the FULL data set.
The grey bands correspond to the 1-$\sigma$ error range, computed in the bootstrap framework. 
For each values of  $E_\gamma$ and $\theta_{\text{lab}}$, we calculate the differential cross section $d\sigma/d\Omega$  as function of the best values of $\astat$ and $\bstat$ obtained at every bootstrap cycle. 
We then have $N=10000$ values for $d\sigma/d\Omega$, from which we can reconstruct its probability distribution and the $68\%$ confidence level range.

	\begin{figure}
  \centering
  \begin{subfigure}[b]{0.45\textwidth}
    \includegraphics[scale=.65]{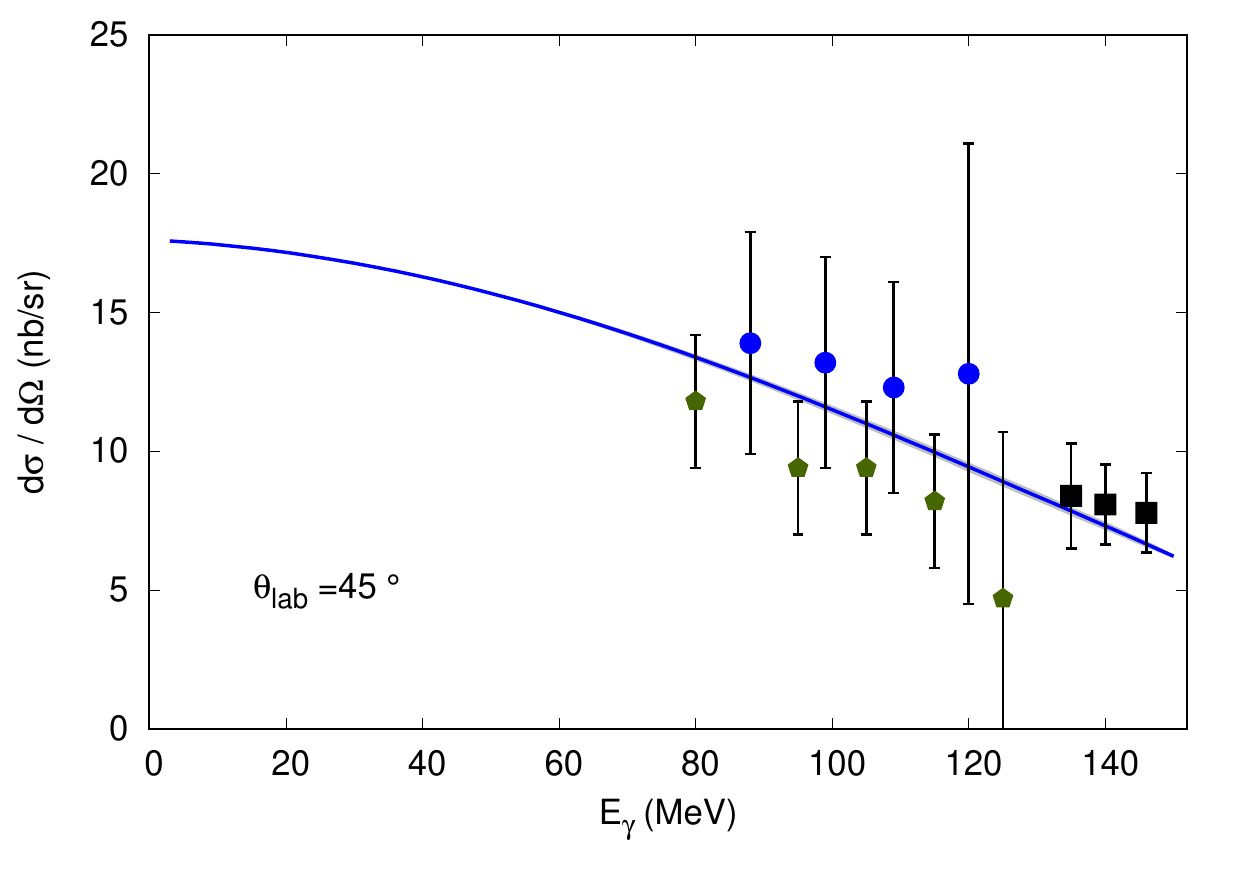}
    \label{fig:cross45}
  \end{subfigure}
  ~ 
  \begin{subfigure}[b]{0.45\textwidth}
    \includegraphics[scale=.65]{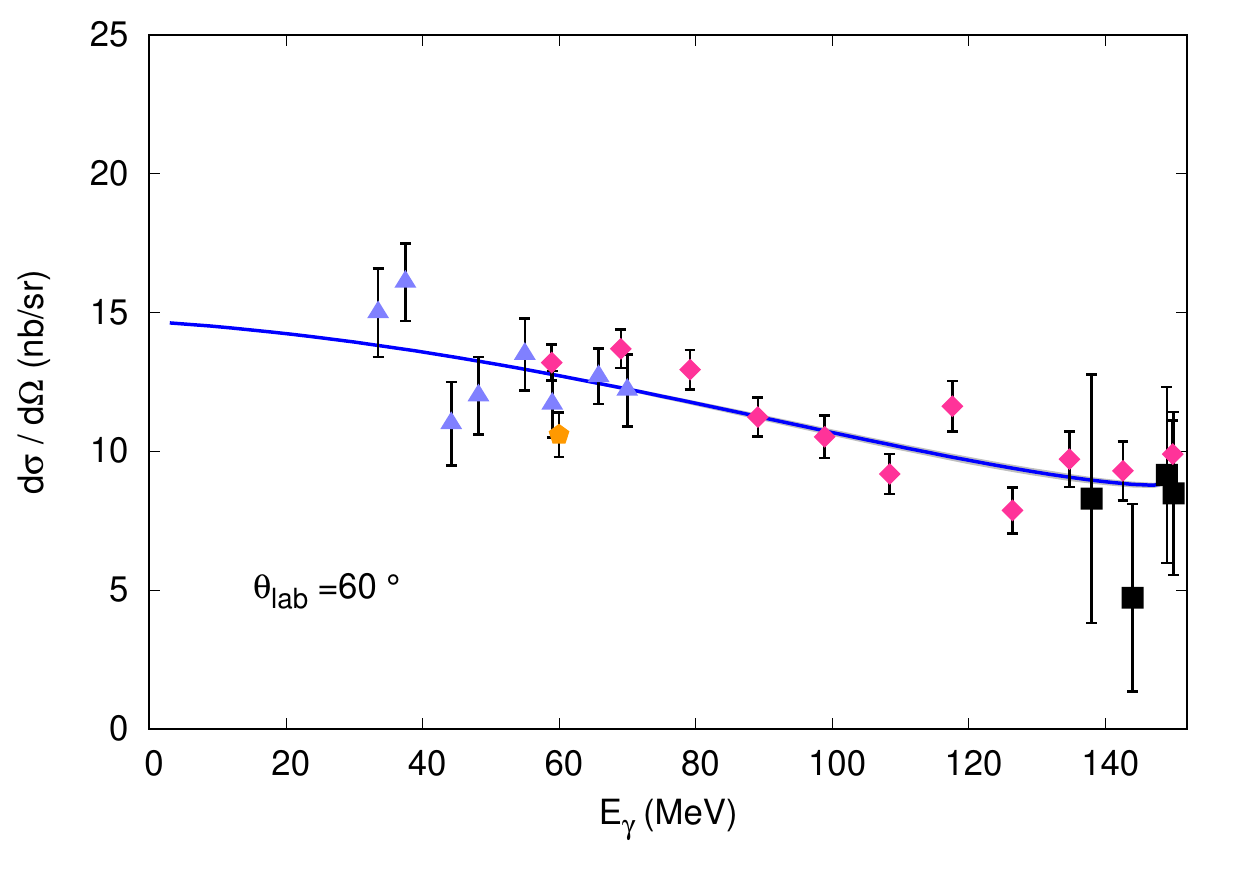}
    \label{fig:cross60}
  \end{subfigure}\\
    \begin{subfigure}[b]{0.45\textwidth}
    \includegraphics[scale=.65]{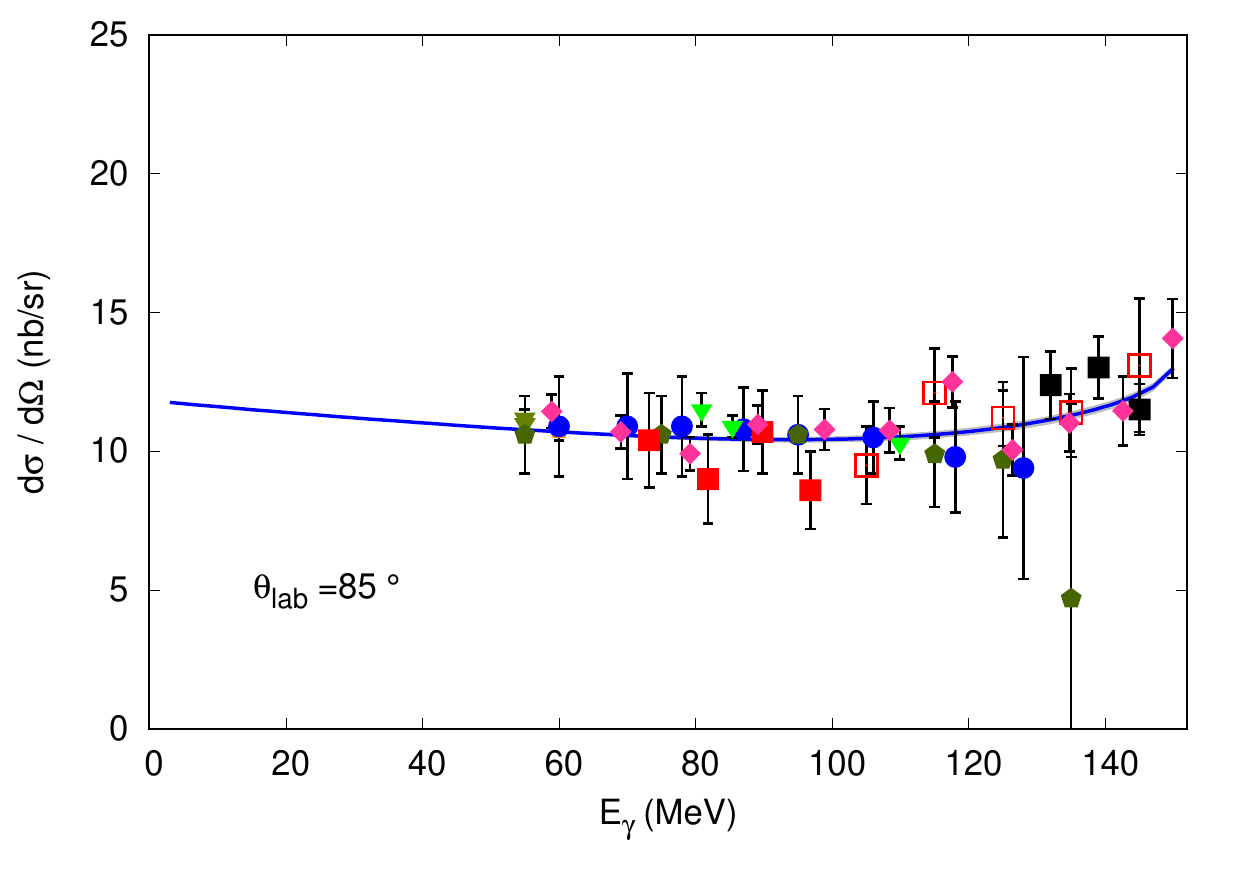}
    \label{fig:cross85}
  \end{subfigure}
  ~ 
  \begin{subfigure}[b]{0.45\textwidth}
    \includegraphics[scale=.65]{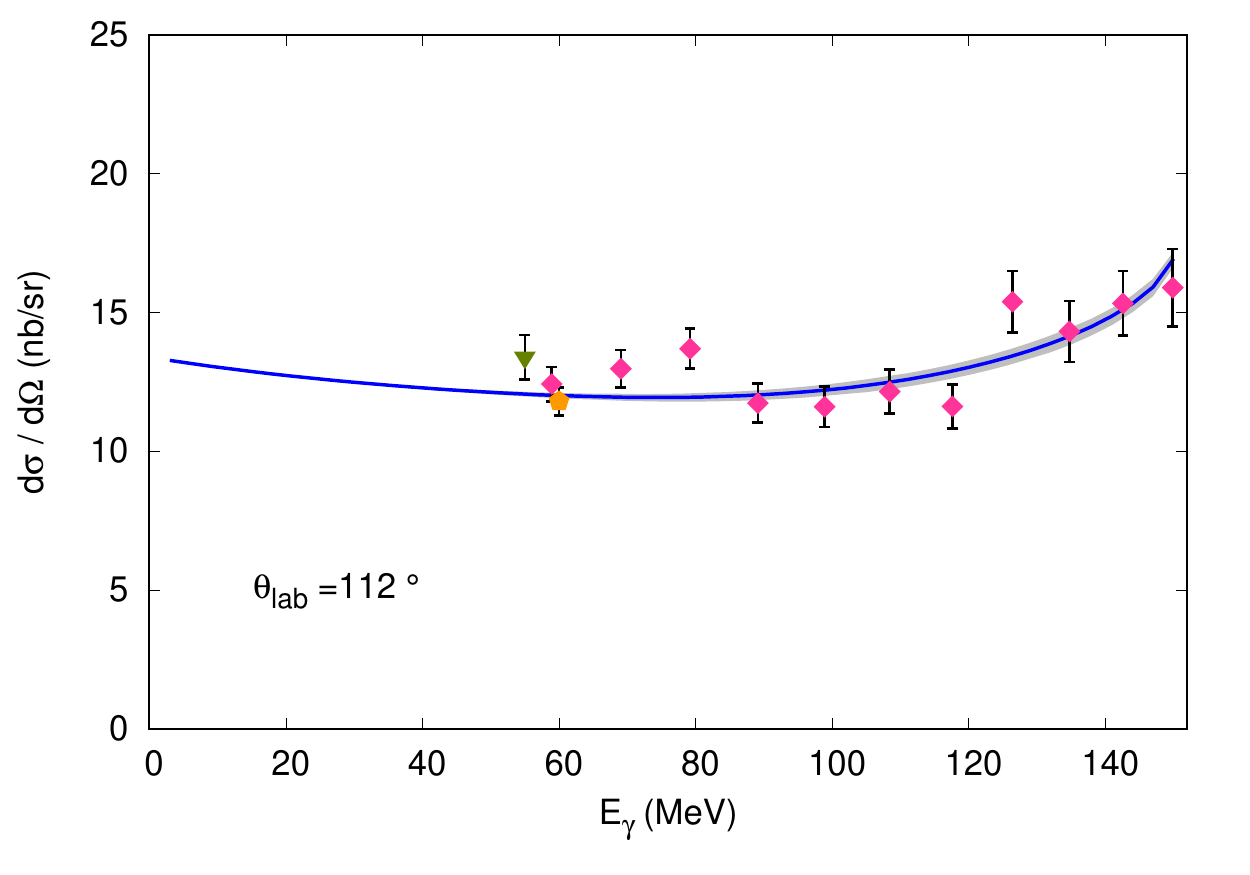}
    \label{fig:cross112}
  \end{subfigure}\\
    \begin{subfigure}[b]{0.45\textwidth}
    \includegraphics[scale=.65]{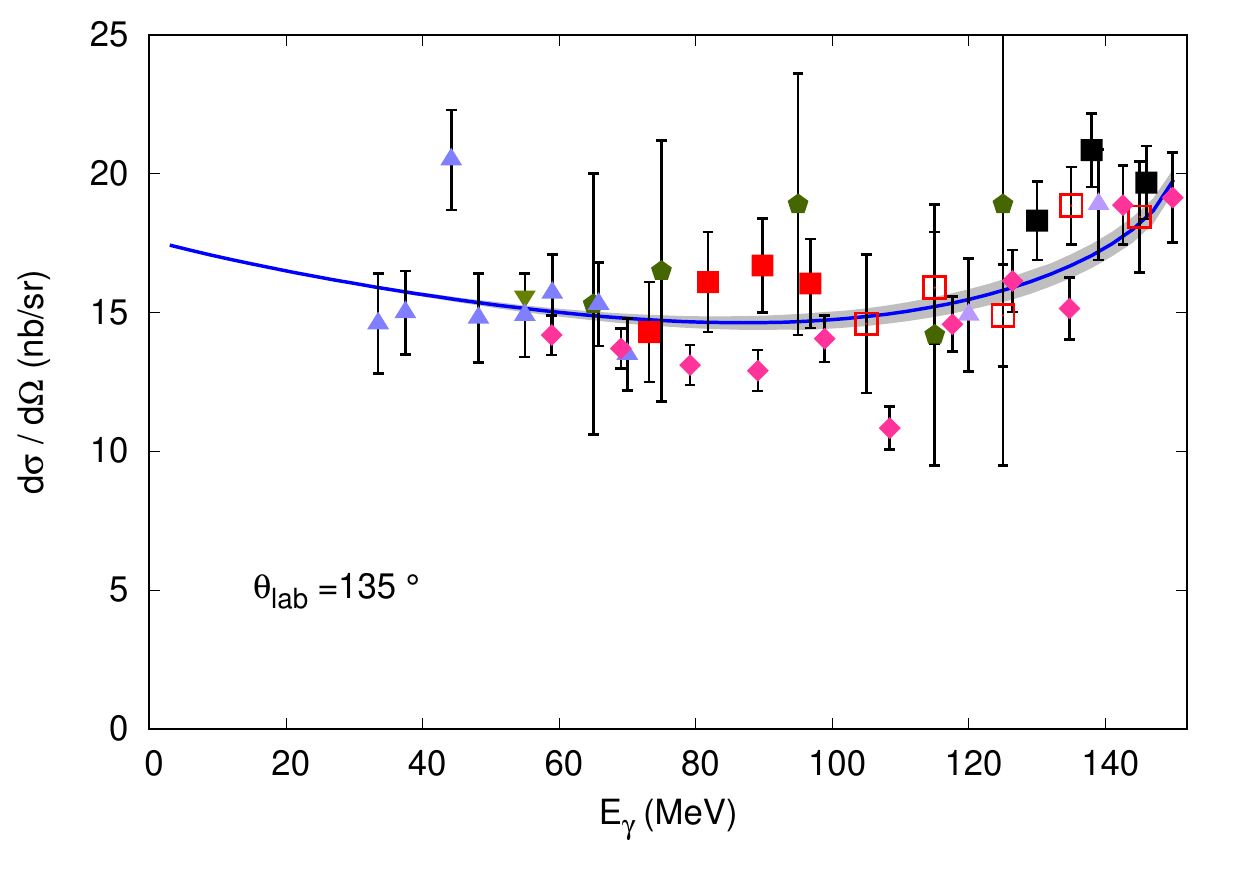}
    \label{fig:cross135}
  \end{subfigure}
  ~ 
  \begin{subfigure}[b]{0.45\textwidth}
    \includegraphics[scale=.65]{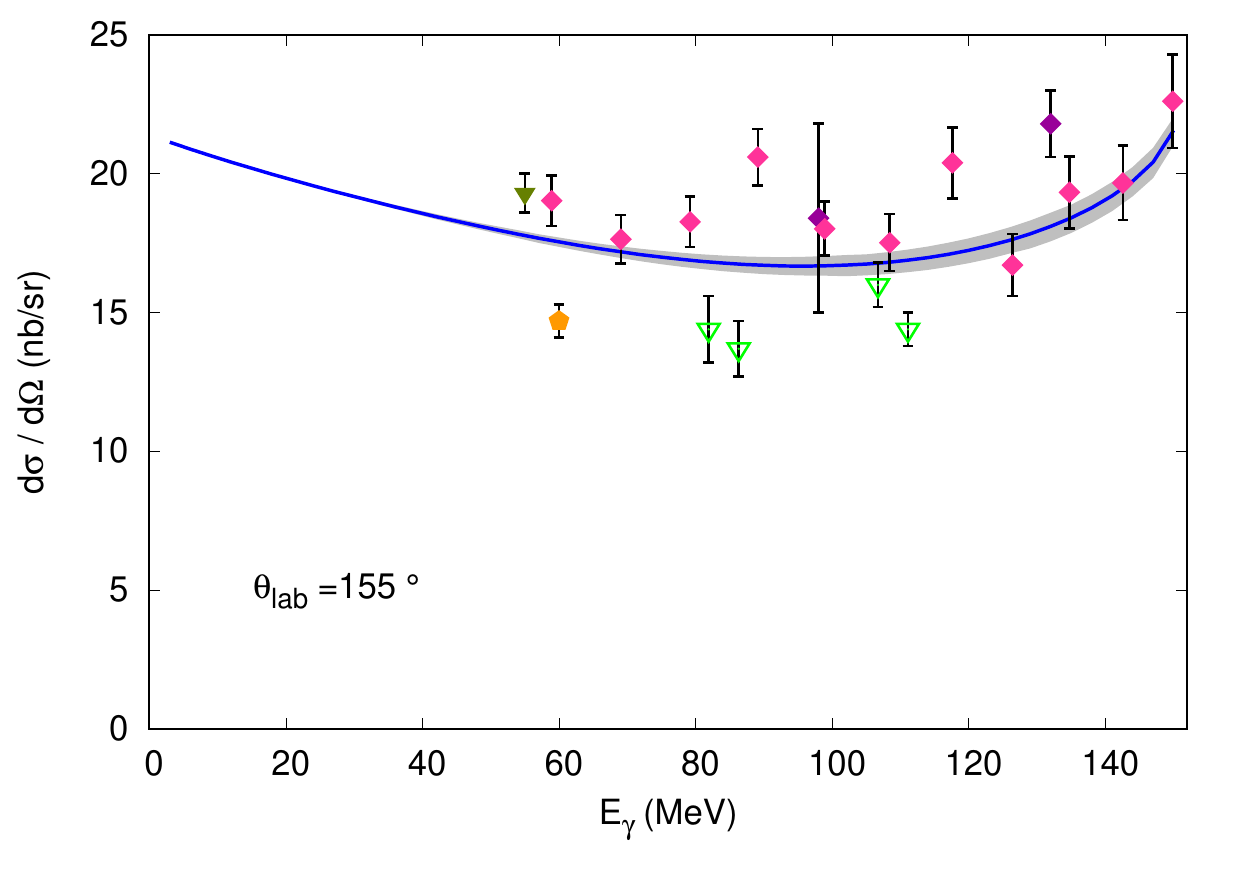}
    \label{fig:cross155}
  \end{subfigure}  
  \caption{The RCS differential cross section (blue line), evaluated with the scalar dipole polarizabilities of \eqr{eq:summary} and the 
experimental values of Ref.~\cite{Martel:2014pba} for the leading-order spin polarizabilities, as function of the lab photon energy ($E_\gamma$) and lab scattering angle ($\theta_{\text{lab}}$).
The gray bands correspond to the 1-$\sigma$ error band obtained in the bootstrap framework (see text for more detail).
The experimental data are from the FULL data set, with the labels reported in \tref{tab:data_summary} of App.~\ref{sec:data}. In the last figure for $\theta_{\text{lab}} =155^\circ$,we also show the two data points at $\theta_{\text{lab}} =180^\circ$ of Ref.~\cite{Zieger:1992jq} .}
\label{fig:cross}
\end{figure}

\begin{figure}[h!]
  \centering
    \includegraphics[scale=0.4]{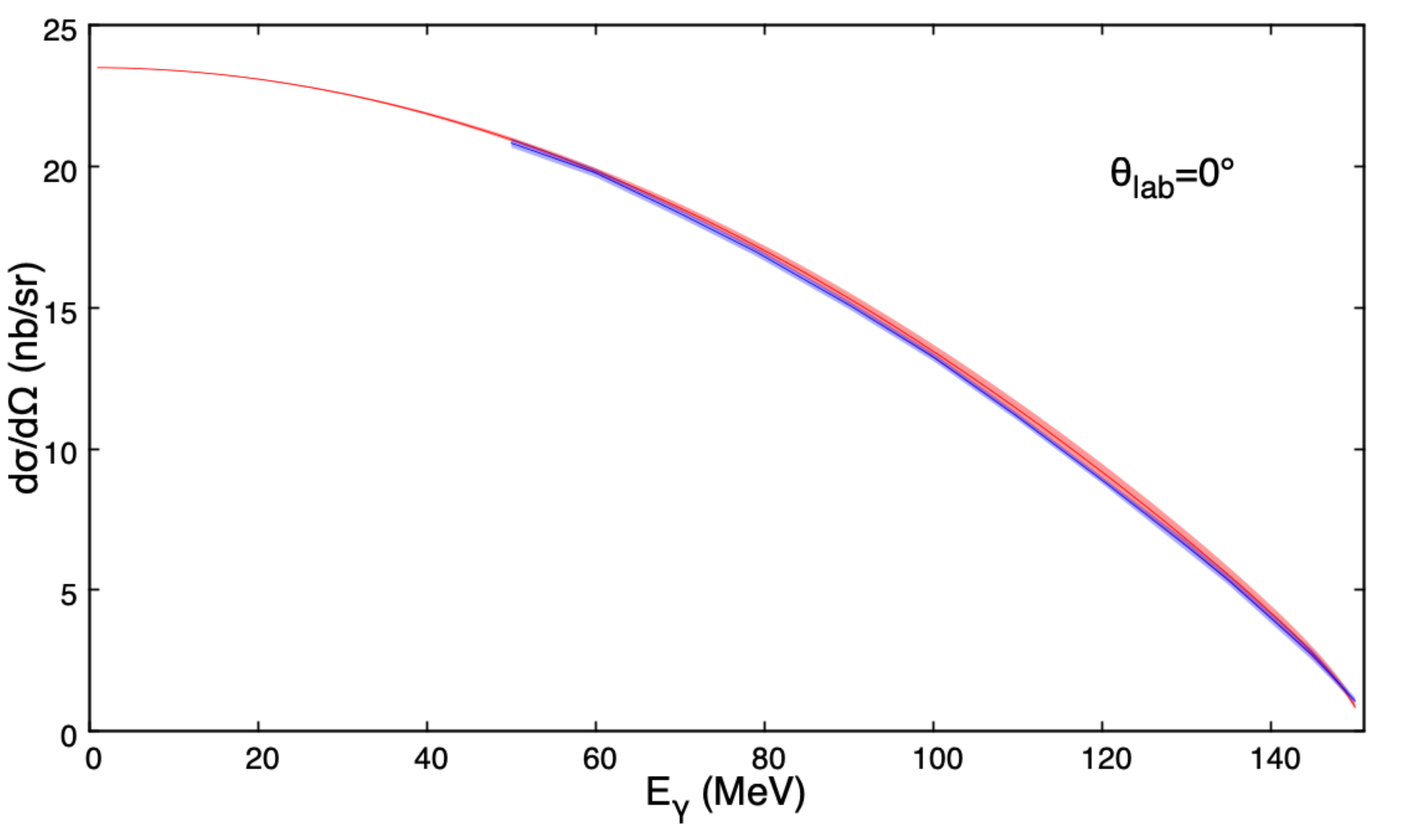}
    \caption{Results for the differential cross section at forward angle as function of the photon lab energy, obtained from the empirical amplitudes of Refs.~\cite{Gryniuk:2016gnm,Gryniuk:2015eza} (blue band) and our analysis (red band).\label{fig-cross-forward}}
      \end{figure}

In Fig.~\ref{fig-cross-forward}, we also show the differential cross section at forward angle $\theta_{\mathrm{lab}}=0^o$ from our analysis (red band) in comparison with the results obtained
with the empirical forward RCS amplitudes of Refs.~\cite{Gryniuk:2016gnm,Gryniuk:2015eza} (blue band). The last ones are evaluated from dispersive sum rules, using as input the total photoabsorption cross sections fitted to the available experimental data. In particular, we used the empirical amplitudes from the fit I of Refs.~\cite{Gryniuk:2016gnm,Gryniuk:2015eza}, that are tabulated from $E_\gamma=50$ MeV and correspond to $\alpha_{E1}+\beta_{M1}=14.29 \pm 0.27$ and $\gamma_0 = -0.929 \pm 0.105$. We observe a remarkable agreement between the two analysis. 

\section{Conclusions}
\label{sec:conclusions}
We performed a fit of the electric $\astat$ and magnetic $\bstat$
polarizabilities to the proton RCS unpolarized cross section data below
pion-production threshold, using subtracted fixed-$t$ DRs
and a bootstrap-based statistical analysis.
Within the subtracted DR formalism, all the leading-order static
polarizabilities enter as subtraction constants to be fitted to the data.
However, due to the limited statistic of the RCS data, a simultaneous fit
of all of them is not achievable at the moment.
We then have restricted ourselves to fit the sets
$\{\astat,\bstat\}$ or
$\{\astat,\bstat,\gamma_\pi\}$, which mainly affect
 the unpolarized RCS cross section below pion-production threshold.
 The remaining spin polarizabilities have been fixed to the available
experimental
information~\cite{Martel:2014pba,Ahrens:2001qt,Dutz:2003mm,Schumacher:2005an}.
 Furthermore, we consider different fit conditions, switching on/off the
systematic errors and with/without the constraint of the Baldin's sum rule
for the polarizability sum $\astat+\bstat$.

We summarized the main features of the parametric-bootstrap method, in
particular the advantages of taking into account
 both the effect of the systematic errors of the experimental data and the
 propagation of the statistical errors of the 
polarizability values not treated as free parameters in the fit procedure.
  
 We showed that the inclusion of the sources of systematic errors in the
 data analysis changes significantly the 
 expected theoretical probability distribution of
the final $\hat \chi^2/d.o.f.$ variable
 and we were able to give realistic p-values for every fitting condition.
 We also presented a critical discussion of the data set consistency.
 We showed some simple but meaningful tests, which
led us to conclude that there are no strong motivations for the exclusion
of any data point from the global RCS data set below pion-production
threshold. Even if
we observed
sizeable deviations between our fit model and two data subsets,  there is not a  clearly identified source of possible experimental problems  
for these data. Therefore, instead of excluding them from the fit, we discussed the possibility to handle them with a suitable rescaling factor of the statistical error bar.
Also the bootstrap fitting technique showed to be useful in these cases to check if the inclusion of  systematic errors in the fitting procedure improves the  
  significance of the fit results.
  
The bootstrap fit using fixed-$t$ subtracted DRs and the global RCS data set below pion-production threshold
yields
 $\astat = (12.03^{+0.48}_{-0.54})\times 10^{-4} \text{fm}^3$ and $\bstat = (1.77^{+0.52}_{-0.54})\times 10^{-4} \text{fm}^3$, with p-value $= 12\%$.
 The results are in agreement with previous analysis obtained with different variants of DRs and the traditional $\chi^2$ fitting procedure. They differ 
 from the extractions using the $\chi$PT frameworks, even if they are compatible within the $2\sigma$ range.
 This discrepancy
 can be traced back  to the different data
sets used in the analyses and, partially, also to the different
theoretical estimates of the higher-order contributions beyond the scalar dipole
polarizabilities
to the RCS cross section.

Future measurements planned by the A2 collaboration at MAMI below pion-production 
threshold \cite{A2-epja, A2-Downie}
hold the promise to improve the accuracy and the statistic of the available data set
and will help to extract  with better precision the values of the proton scalar dipole polarizabilities.

\section{Acknowledgments}
We are grateful to V. Bertone and A. Rotondi 
for a careful reading of the manuscript and useful comments.
We thank D. Phillips for stimulating  discussions  and  useful suggestions on the fitting procedure,
and H. Griesshammer, J. Mc Govern and V. Lensky for the help on the correct representation of the results of $\chi$PT.  

\begin{appendix}
\section{Data sets}
	\label{sec:data}
	In \tref{tab:data_summary}, we list all the available data sets for RCS in the energy range below pion production threshold ($\sim 150 $ MeV in lab frame). For the sets \cite{Oxley:1958zz,Hyman:1959zz,GOLDANSKY1960473} and \cite{Pugh:1957zz}, we use the Baranov data-selection~\cite{Baranov:2001jv}.
	Furthermore, as done also in Ref.~\cite{McGovern:2012ew,Griesshammer:2012we}, we discard the data from Table I in the Hallin paper~\cite{Hallin:1993ft}, because it is not clear if they are really independent from the data given in Table II of the same work.
The data sets used in our analysis are:
	\beitem
\item FULL, which includes all the available data sets below pion-production threshold listed in \tref{tab:data_summary}, for a total of 150 data points.
\item SELECTED, which is based on the data selection proposed in Ref.~\cite{McGovern:2012ew,Griesshammer:2012we}, corresponding to the FULL data set except for the data from Ref.~\cite{Oxley:1958zz,Bernardini:1960wya,Baranov:1975ju}, a single point ($\theta_{\text{lab}} = 133^\circ$, $E_\gamma = 108$ MeV) from Ref.~\cite{OlmosdeLeon:2001zn} and a single point ($\theta_{\text{lab}} =135^\circ$, $E_\gamma =44$ MeV) from Ref.~\cite{Federspiel:1991yd}, for a total of 137 data points.
\item TAPS, which is the most comprehensive available subset with 55 data points below pion-production threshold~\cite{OlmosdeLeon:2001zn}.
	\enitem
The sets 6 and 7 from Ref.~\cite{Baranov:1974ec,Baranov:1975ju} are from the same experimental measurements, but they differ for the values of the systematic errors. The same for the sets 11 and 12 from Ref.~\cite{MacGibbon:1995in}.
	\betab[h]
\begin{tabular}{|c|c|c|c|c|c|c|}
\hline
&&&&&&\\
set label & Ref. &first author&points number& $\theta_{\text{lab}}$ ($^\circ$) & $E_\gamma$ (MeV) & symbol\\
\hline
1 & \cite{Oxley:1958zz} & Oxley & 4 & $70-150$ &$\simeq 60$ & \oxl \\
2 & \cite{Hyman:1959zz} & Hyman&12 & $50,90$ & $55-95$& \hym\\
3 & \cite{GOLDANSKY1960473} &Goldansky&5 &$75-150$ & $55-80$& \gold\\
4 & \cite{Bernardini:1960wya} & Bernardini &2 & $\simeq 135$ & $\simeq 140$ & \ber \\
5 & \cite{Pugh:1957zz} & Pugh&16 & $50-135$ & $40-120$ & \pugh \\
6 & \cite{Baranov:1974ec,Baranov:1975ju} &Baranov &3 & $90,150$ & $80-110$ &\baranov \\
7 & \cite{Baranov:1974ec,Baranov:1975ju} &Baranov &4 & $90,150$ & $80-110$ &\baranovbis \\
8 & \cite{Federspiel:1991yd} & Federspiel&16 & $60,135$ & $30-90$ &\fed\\
9 & \cite{Zieger:1992jq} & Zieger&2 & $180$ & $100,130$ &\zie\\
10 & \cite{Hallin:1993ft} & Hallin&13 & $45-135$ & $130-150$ &\hal\\
11 & \cite{MacGibbon:1995in} & MacGibbon&8 & $90,135$ & $95-145$ &\mcg\\
12 & \cite{MacGibbon:1995in} & MacGibbon&10 & $90,135$ & $95-145$ &\mcgbis\\
13 & \cite{OlmosdeLeon:2001zn}& Olmos de Leon &55 & $60-155$ & $60-150$ &\taps\\
\hline
\end{tabular}
\caption{Angular and energy coverage of the available experimental data on unpolarized cross section for proton RCS.}\label{tab:data_summary}
	\entab
\section{Correlation coefficients among fit parameters}
\label{sec:correlations}
In the bootstrap framework, the correlation coefficients $\rho$ among the fit parameters are obtained from the reconstructed probability distribution in the parameters space.
In Table~\ref{tab:correlations}, we list these coefficients for all the different fitting conditions used in this work.

 In the Baldin-constrained fits, we do not obtain $\rho_{\astat-\bstat} = -1$, due to the fact that $\astat + \bstat$ is not fixed to its central value, but is sampled within its uncertainty
 with a Gaussian distribution, as explained in \secref{sec:errprop}. This  behavior was already observed in the extraction of the scalar dipole dynamical polarizabilities in Ref.~\cite{Pasquini:2017ehj}. 
We also note a large and negative (positive) correlation between $\gamma_\pi$ and $\bstat$ ($\astat$).
This behavior is mainly a
consequence of low sensitivity of the existing data
to the $\gamma_\pi$ polarizability.
\betab[t!]
	\renewcommand{\arraystretch}{1.7}
\begin{tabular}{|c|c|c|c|c|}
\hline
\multicolumn{4}{|c|}{FULL data set}\\
\hline
\hline
fit conditions & $\rho_{\astat-\bstat}$ & $\rho_{\astat-\gamma_\pi}$ & $\rho_{\bstat-\gamma_\pi}$\\
\hline
Fit1 & $-0.64$ & $--$ & $--$\\
Fit $1^\prime$ & $-0.72$ & $--$ & $--$\\ 
Fit 2 & $0.59$ & $--$ & $--$\\ 
Fit $2^\prime$ & $0.52$ & $--$ & $--$\\ 
Fit 3 & $-0.84$ & $0.86$ & $-0.88$\\ 
Fit $3^\prime$ & $-0.87$ & $0.84$ & $-0.86$\\ 
\hline
\hline
\multicolumn{4}{|c|}{TAPS data set}\\
\hline
\hline
fit conditions & $\rho_{\astat-\bstat}$ & $\rho_{\astat-\gamma_\pi}$ & $\rho_{\bstat-\gamma_\pi}$\\
\hline
Fit 1 & $-0.74$ & $--$ & $--$\\ 
Fit $1^\prime$ & $-0.74$ & $--$ & $--$\\
Fit 2 & $0.47$ & $--$ & $--$\\ 
Fit $2^\prime$ &$0.23$ & $--$ & $--$\\ 
Fit 3 & $-0.85$ & $0.82$ & $-0.84$\\ 
Fit $3^\prime$ & $-0.86$ & $0.81$ & $-0.83$\\
\hline
\end{tabular}
\caption{Correlation coefficients $\rho$ among the fit parameters in the different fitting conditions described in Sect.~\ref{subsect:results}. 
The columns 2-4 correspond, from the left to the right, to the correlation coefficients between $\astat$ and $\bstat$, $\astat$ and $\gamma_\pi$, $\bstat$ and $\gamma_\pi$.}\label{tab:correlations}
	\entab

\end{appendix}

\clearpage
\input{draft_revised_final.bbl}
\end{document}

%% file: draft_revised_final.bbl
%

%% file: draft_revised_final.bbl
\begin{thebibliography}{58}%
\makeatletter
\providecommand \@ifxundefined [1]{%
 \@ifx{#1\undefined}
}%
\providecommand \@ifnum [1]{%
 \ifnum #1\expandafter \@firstoftwo
 \else \expandafter \@secondoftwo
 \fi
}%
\providecommand \@ifx [1]{%
 \ifx #1\expandafter \@firstoftwo
 \else \expandafter \@secondoftwo
 \fi
}%
\providecommand \natexlab [1]{#1}%
\providecommand \enquote  [1]{``#1''}%
\providecommand \bibnamefont  [1]{#1}%
\providecommand \bibfnamefont [1]{#1}%
\providecommand \citenamefont [1]{#1}%
\providecommand \href@noop [0]{\@secondoftwo}%
\providecommand \href [0]{\begingroup \@sanitize@url \@href}%
\providecommand \@href[1]{\@@startlink{#1}\@@href}%
\providecommand \@@href[1]{\endgroup#1\@@endlink}%
\providecommand \@sanitize@url [0]{\catcode `\\12\catcode `\$12\catcode
  `\&12\catcode `\#12\catcode `\^12\catcode `\_12\catcode `\%12\relax}%
\providecommand \@@startlink[1]{}%
\providecommand \@@endlink[0]{}%
\providecommand \url  [0]{\begingroup\@sanitize@url \@url }%
\providecommand \@url [1]{\endgroup\@href {#1}{\urlprefix }}%
\providecommand \urlprefix  [0]{URL }%
\providecommand \Eprint [0]{\href }%
\providecommand \doibase [0]{http://dx.doi.org/}%
\providecommand \selectlanguage [0]{\@gobble}%
\providecommand \bibinfo  [0]{\@secondoftwo}%
\providecommand \bibfield  [0]{\@secondoftwo}%
\providecommand \translation [1]{[#1]}%
\providecommand \BibitemOpen [0]{}%
\providecommand \bibitemStop [0]{}%
\providecommand \bibitemNoStop [0]{.\EOS\space}%
\providecommand \EOS [0]{\spacefactor3000\relax}%
\providecommand \BibitemShut  [1]{\csname bibitem#1\endcsname}%
\let\auto@bib@innerbib\@empty
\bibitem [{\citenamefont {L'vov}\ \emph {et~al.}(1997)\citenamefont {L'vov},
  \citenamefont {Petrun'kin},\ and\ \citenamefont {Schumacher}}]{Lvov:1996rmi}%
  \BibitemOpen
  \bibfield  {author} {\bibinfo {author} {\bibfnamefont {A.~I.}\ \bibnamefont
  {L'vov}}, \bibinfo {author} {\bibfnamefont {V.~A.}\ \bibnamefont
  {Petrun'kin}}, \ and\ \bibinfo {author} {\bibfnamefont {M.}~\bibnamefont
  {Schumacher}},\ }\href {\doibase 10.1103/PhysRevC.55.359} {\bibfield
  {journal} {\bibinfo  {journal} {Phys. Rev. C}\ }\textbf {\bibinfo {volume}
  {55}},\ \bibinfo {pages} {359} (\bibinfo {year} {1997})}\BibitemShut
  {NoStop}%
\bibitem [{\citenamefont {Babusci}\ \emph {et~al.}(1998)\citenamefont
  {Babusci}, \citenamefont {Giordano}, \citenamefont {L'vov}, \citenamefont
  {Matone},\ and\ \citenamefont {Nathan}}]{Babusci:1998ww}%
  \BibitemOpen
  \bibfield  {author} {\bibinfo {author} {\bibfnamefont {D.}~\bibnamefont
  {Babusci}}, \bibinfo {author} {\bibfnamefont {G.}~\bibnamefont {Giordano}},
  \bibinfo {author} {\bibfnamefont {A.}~\bibnamefont {L'vov}}, \bibinfo
  {author} {\bibfnamefont {G.}~\bibnamefont {Matone}}, \ and\ \bibinfo {author}
  {\bibfnamefont {A.}~\bibnamefont {Nathan}},\ }\href {\doibase
  10.1103/PhysRevC.58.1013} {\bibfield  {journal} {\bibinfo  {journal} {Phys.
  Rev. C}\ }\textbf {\bibinfo {volume} {58}},\ \bibinfo {pages} {1013}
  (\bibinfo {year} {1998})},\ \Eprint {http://arxiv.org/abs/hep-ph/9803347}
  {arXiv:hep-ph/9803347 [hep-ph]} \BibitemShut {NoStop}%
\bibitem [{\citenamefont {Schumacher}(2005)}]{Schumacher:2005an}%
  \BibitemOpen
  \bibfield  {author} {\bibinfo {author} {\bibfnamefont {M.}~\bibnamefont
  {Schumacher}},\ }\href {\doibase 10.1016/j.ppnp.2005.01.033} {\bibfield
  {journal} {\bibinfo  {journal} {Prog. Part. Nucl. Phys.}\ }\textbf {\bibinfo
  {volume} {55}},\ \bibinfo {pages} {567} (\bibinfo {year} {2005})},\ \Eprint
  {http://arxiv.org/abs/hep-ph/0501167} {arXiv:hep-ph/0501167 [hep-ph]}
  \BibitemShut {NoStop}%
\bibitem [{\citenamefont {Drechsel}\ \emph {et~al.}(1999)\citenamefont
  {Drechsel}, \citenamefont {Gorchtein}, \citenamefont {Pasquini},\ and\
  \citenamefont {Vanderhaeghen}}]{Drechsel:1999rf}%
  \BibitemOpen
  \bibfield  {author} {\bibinfo {author} {\bibfnamefont {D.}~\bibnamefont
  {Drechsel}}, \bibinfo {author} {\bibfnamefont {M.}~\bibnamefont {Gorchtein}},
  \bibinfo {author} {\bibfnamefont {B.}~\bibnamefont {Pasquini}}, \ and\
  \bibinfo {author} {\bibfnamefont {M.}~\bibnamefont {Vanderhaeghen}},\ }\href
  {\doibase 10.1103/PhysRevC.61.015204} {\bibfield  {journal} {\bibinfo
  {journal} {Phys. Rev. C}\ }\textbf {\bibinfo {volume} {61}},\ \bibinfo
  {pages} {015204} (\bibinfo {year} {1999})},\ \Eprint
  {http://arxiv.org/abs/hep-ph/9904290} {arXiv:hep-ph/9904290 [hep-ph]}
  \BibitemShut {NoStop}%
\bibitem [{\citenamefont {Holstein}\ \emph {et~al.}(2000)\citenamefont
  {Holstein}, \citenamefont {Drechsel}, \citenamefont {Pasquini},\ and\
  \citenamefont {Vanderhaeghen}}]{Holstein:1999uu}%
  \BibitemOpen
  \bibfield  {author} {\bibinfo {author} {\bibfnamefont {B.~R.}\ \bibnamefont
  {Holstein}}, \bibinfo {author} {\bibfnamefont {D.}~\bibnamefont {Drechsel}},
  \bibinfo {author} {\bibfnamefont {B.}~\bibnamefont {Pasquini}}, \ and\
  \bibinfo {author} {\bibfnamefont {M.}~\bibnamefont {Vanderhaeghen}},\ }\href
  {\doibase 10.1103/PhysRevC.61.034316} {\bibfield  {journal} {\bibinfo
  {journal} {Phys. Rev. C}\ }\textbf {\bibinfo {volume} {61}},\ \bibinfo
  {pages} {034316} (\bibinfo {year} {2000})},\ \Eprint
  {http://arxiv.org/abs/hep-ph/9910427} {arXiv:hep-ph/9910427 [hep-ph]}
  \BibitemShut {NoStop}%
\bibitem [{\citenamefont {Pasquini}\ \emph {et~al.}(2007)\citenamefont
  {Pasquini}, \citenamefont {Drechsel},\ and\ \citenamefont
  {Vanderhaeghen}}]{Pasquini:2007hf}%
  \BibitemOpen
  \bibfield  {author} {\bibinfo {author} {\bibfnamefont {B.}~\bibnamefont
  {Pasquini}}, \bibinfo {author} {\bibfnamefont {D.}~\bibnamefont {Drechsel}},
  \ and\ \bibinfo {author} {\bibfnamefont {M.}~\bibnamefont {Vanderhaeghen}},\
  }\href {\doibase 10.1103/PhysRevC.76.015203} {\bibfield  {journal} {\bibinfo
  {journal} {Phys. Rev. C}\ }\textbf {\bibinfo {volume} {76}},\ \bibinfo
  {pages} {015203} (\bibinfo {year} {2007})},\ \Eprint
  {http://arxiv.org/abs/0705.0282} {arXiv:0705.0282 [hep-ph]} \BibitemShut
  {NoStop}%
\bibitem [{\citenamefont {Drechsel}\ \emph {et~al.}(2003)\citenamefont
  {Drechsel}, \citenamefont {Pasquini},\ and\ \citenamefont
  {Vanderhaeghen}}]{Drechsel:2002ar}%
  \BibitemOpen
  \bibfield  {author} {\bibinfo {author} {\bibfnamefont {D.}~\bibnamefont
  {Drechsel}}, \bibinfo {author} {\bibfnamefont {B.}~\bibnamefont {Pasquini}},
  \ and\ \bibinfo {author} {\bibfnamefont {M.}~\bibnamefont {Vanderhaeghen}},\
  }\href {\doibase 10.1016/S0370-1573(02)00636-1} {\bibfield  {journal}
  {\bibinfo  {journal} {Phys. Rept.}\ }\textbf {\bibinfo {volume} {378}},\
  \bibinfo {pages} {99} (\bibinfo {year} {2003})},\ \Eprint
  {http://arxiv.org/abs/hep-ph/0212124} {arXiv:hep-ph/0212124 [hep-ph]}
  \BibitemShut {NoStop}%
\bibitem [{\citenamefont {Pasquini}\ and\ \citenamefont
  {Vanderhaeghen}(2018)}]{Pasquini:2018wbl}%
  \BibitemOpen
  \bibfield  {author} {\bibinfo {author} {\bibfnamefont {B.}~\bibnamefont
  {Pasquini}}\ and\ \bibinfo {author} {\bibfnamefont {M.}~\bibnamefont
  {Vanderhaeghen}},\ }\href {\doibase 10.1146/annurev-nucl-101917-020843}
  {\bibfield  {journal} {\bibinfo  {journal} {Ann. Rev. Nucl. Part. Sci.}\
  }\textbf {\bibinfo {volume} {68}},\ \bibinfo {pages} {75} (\bibinfo {year}
  {2018})},\ \Eprint {http://arxiv.org/abs/1805.10482} {arXiv:1805.10482
  [hep-ph]} \BibitemShut {NoStop}%
\bibitem [{\citenamefont {Bernard}\ \emph {et~al.}(1995)\citenamefont
  {Bernard}, \citenamefont {Kaiser},\ and\ \citenamefont
  {Meissner}}]{Bernard:1995dp}%
  \BibitemOpen
  \bibfield  {author} {\bibinfo {author} {\bibfnamefont {V.}~\bibnamefont
  {Bernard}}, \bibinfo {author} {\bibfnamefont {N.}~\bibnamefont {Kaiser}}, \
  and\ \bibinfo {author} {\bibfnamefont {U.-G.}\ \bibnamefont {Meissner}},\
  }\href {\doibase 10.1142/S0218301395000092} {\bibfield  {journal} {\bibinfo
  {journal} {Int. J. Mod. Phys.}\ }\textbf {\bibinfo {volume} {E4}},\ \bibinfo
  {pages} {193} (\bibinfo {year} {1995})},\ \Eprint
  {http://arxiv.org/abs/hep-ph/9501384} {arXiv:hep-ph/9501384 [hep-ph]}
  \BibitemShut {NoStop}%
\bibitem [{\citenamefont {Beane}\ \emph {et~al.}(2003)\citenamefont {Beane},
  \citenamefont {Malheiro}, \citenamefont {McGovern}, \citenamefont
  {Phillips},\ and\ \citenamefont {van Kolck}}]{Beane:2002wn}%
  \BibitemOpen
  \bibfield  {author} {\bibinfo {author} {\bibfnamefont {S.~R.}\ \bibnamefont
  {Beane}}, \bibinfo {author} {\bibfnamefont {M.}~\bibnamefont {Malheiro}},
  \bibinfo {author} {\bibfnamefont {J.~A.}\ \bibnamefont {McGovern}}, \bibinfo
  {author} {\bibfnamefont {D.~R.}\ \bibnamefont {Phillips}}, \ and\ \bibinfo
  {author} {\bibfnamefont {U.}~\bibnamefont {van Kolck}},\ }\href {\doibase
  10.1016/j.physletb.2003.06.040, 10.1016/j.physletb.2004.12.069} {\bibfield
  {journal} {\bibinfo  {journal} {Phys. Lett.}\ }\textbf {\bibinfo {volume}
  {B567}},\ \bibinfo {pages} {200} (\bibinfo {year} {2003})},\ \bibinfo {note}
  {[Erratum: Phys. Lett.B607,320(2005)]},\ \Eprint
  {http://arxiv.org/abs/nucl-th/0209002} {arXiv:nucl-th/0209002 [nucl-th]}
  \BibitemShut {NoStop}%
\bibitem [{\citenamefont {McGovern}\ \emph {et~al.}(2013)\citenamefont
  {McGovern}, \citenamefont {Phillips},\ and\ \citenamefont
  {Griesshammer}}]{McGovern:2012ew}%
  \BibitemOpen
  \bibfield  {author} {\bibinfo {author} {\bibfnamefont {J.~A.}\ \bibnamefont
  {McGovern}}, \bibinfo {author} {\bibfnamefont {D.~R.}\ \bibnamefont
  {Phillips}}, \ and\ \bibinfo {author} {\bibfnamefont {H.~W.}\ \bibnamefont
  {Griesshammer}},\ }\href {\doibase 10.1140/epja/i2013-13012-1} {\bibfield
  {journal} {\bibinfo  {journal} {Eur. Phys. J. A}\ }\textbf {\bibinfo {volume}
  {49}},\ \bibinfo {pages} {12} (\bibinfo {year} {2013})},\ \Eprint
  {http://arxiv.org/abs/1210.4104} {arXiv:1210.4104 [nucl-th]} \BibitemShut
  {NoStop}%
\bibitem [{\citenamefont {Lensky}\ \emph {et~al.}(2015)\citenamefont {Lensky},
  \citenamefont {McGovern},\ and\ \citenamefont {Pascalutsa}}]{Lensky:2015awa}%
  \BibitemOpen
  \bibfield  {author} {\bibinfo {author} {\bibfnamefont {V.}~\bibnamefont
  {Lensky}}, \bibinfo {author} {\bibfnamefont {J.}~\bibnamefont {McGovern}}, \
  and\ \bibinfo {author} {\bibfnamefont {V.}~\bibnamefont {Pascalutsa}},\
  }\href {\doibase 10.1140/epjc/s10052-015-3791-0} {\bibfield  {journal}
  {\bibinfo  {journal} {Eur. Phys. J. C}\ }\textbf {\bibinfo {volume} {75}},\
  \bibinfo {pages} {604} (\bibinfo {year} {2015})},\ \Eprint
  {http://arxiv.org/abs/1510.02794} {arXiv:1510.02794 [hep-ph]} \BibitemShut
  {NoStop}%
\bibitem [{\citenamefont {Lensky}\ and\ \citenamefont
  {Pascalutsa}(2010)}]{Lensky:2009uv}%
  \BibitemOpen
  \bibfield  {author} {\bibinfo {author} {\bibfnamefont {V.}~\bibnamefont
  {Lensky}}\ and\ \bibinfo {author} {\bibfnamefont {V.}~\bibnamefont
  {Pascalutsa}},\ }\href {\doibase 10.1140/epjc/s10052-009-1183-z} {\bibfield
  {journal} {\bibinfo  {journal} {Eur. Phys. J.}\ }\textbf {\bibinfo {volume}
  {C65}},\ \bibinfo {pages} {195} (\bibinfo {year} {2010})},\ \Eprint
  {http://arxiv.org/abs/0907.0451} {arXiv:0907.0451 [hep-ph]} \BibitemShut
  {NoStop}%
\bibitem [{\citenamefont {Kondratyuk}\ and\ \citenamefont
  {Scholten}(2001)}]{Kondratyuk:2001qu}%
  \BibitemOpen
  \bibfield  {author} {\bibinfo {author} {\bibfnamefont {S.}~\bibnamefont
  {Kondratyuk}}\ and\ \bibinfo {author} {\bibfnamefont {O.}~\bibnamefont
  {Scholten}},\ }\href {\doibase 10.1103/PhysRevC.64.024005} {\bibfield
  {journal} {\bibinfo  {journal} {Phys. Rev.}\ }\textbf {\bibinfo {volume}
  {C64}},\ \bibinfo {pages} {024005} (\bibinfo {year} {2001})},\ \Eprint
  {http://arxiv.org/abs/nucl-th/0103006} {arXiv:nucl-th/0103006 [nucl-th]}
  \BibitemShut {NoStop}%
\bibitem [{\citenamefont {Gasparyan}\ \emph {et~al.}(2011)\citenamefont
  {Gasparyan}, \citenamefont {Lutz},\ and\ \citenamefont
  {Pasquini}}]{Gasparyan:2011yw}%
  \BibitemOpen
  \bibfield  {author} {\bibinfo {author} {\bibfnamefont {A.~M.}\ \bibnamefont
  {Gasparyan}}, \bibinfo {author} {\bibfnamefont {M.~F.~M.}\ \bibnamefont
  {Lutz}}, \ and\ \bibinfo {author} {\bibfnamefont {B.}~\bibnamefont
  {Pasquini}},\ }\href {\doibase 10.1016/j.nuclphysa.2011.07.005} {\bibfield
  {journal} {\bibinfo  {journal} {Nucl. Phys.}\ }\textbf {\bibinfo {volume}
  {A866}},\ \bibinfo {pages} {79} (\bibinfo {year} {2011})},\ \Eprint
  {http://arxiv.org/abs/1102.3375} {arXiv:1102.3375 [hep-ph]} \BibitemShut
  {NoStop}%
\bibitem [{\citenamefont {Pasquini}\ \emph {et~al.}(2018)\citenamefont
  {Pasquini}, \citenamefont {Pedroni},\ and\ \citenamefont
  {Sconfietti}}]{Pasquini:2017ehj}%
  \BibitemOpen
  \bibfield  {author} {\bibinfo {author} {\bibfnamefont {B.}~\bibnamefont
  {Pasquini}}, \bibinfo {author} {\bibfnamefont {P.}~\bibnamefont {Pedroni}}, \
  and\ \bibinfo {author} {\bibfnamefont {S.}~\bibnamefont {Sconfietti}},\
  }\href {\doibase 10.1103/PhysRevC.98.015204} {\bibfield  {journal} {\bibinfo
  {journal} {Phys. Rev.}\ }\textbf {\bibinfo {volume} {C98}},\ \bibinfo {pages}
  {015204} (\bibinfo {year} {2018})},\ \Eprint
  {http://arxiv.org/abs/1711.07401} {arXiv:1711.07401 [hep-ph]} \BibitemShut
  {NoStop}%
\bibitem [{\citenamefont {Griesshammer}\ and\ \citenamefont
  {Hemmert}(2002)}]{Griesshammer:2001uw}%
  \BibitemOpen
  \bibfield  {author} {\bibinfo {author} {\bibfnamefont {H.~W.}\ \bibnamefont
  {Griesshammer}}\ and\ \bibinfo {author} {\bibfnamefont {T.~R.}\ \bibnamefont
  {Hemmert}},\ }\href {\doibase 10.1103/PhysRevC.65.045207} {\bibfield
  {journal} {\bibinfo  {journal} {Phys. Rev. C}\ }\textbf {\bibinfo {volume}
  {65}},\ \bibinfo {pages} {045207} (\bibinfo {year} {2002})},\ \Eprint
  {http://arxiv.org/abs/nucl-th/0110006} {arXiv:nucl-th/0110006 [nucl-th]}
  \BibitemShut {NoStop}%
\bibitem [{\citenamefont {Hildebrandt}\ \emph {et~al.}(2004)\citenamefont
  {Hildebrandt}, \citenamefont {Griesshammer}, \citenamefont {Hemmert},\ and\
  \citenamefont {Pasquini}}]{Hildebrandt:2003fm}%
  \BibitemOpen
  \bibfield  {author} {\bibinfo {author} {\bibfnamefont {R.~P.}\ \bibnamefont
  {Hildebrandt}}, \bibinfo {author} {\bibfnamefont {H.~W.}\ \bibnamefont
  {Griesshammer}}, \bibinfo {author} {\bibfnamefont {T.~R.}\ \bibnamefont
  {Hemmert}}, \ and\ \bibinfo {author} {\bibfnamefont {B.}~\bibnamefont
  {Pasquini}},\ }\href {\doibase 10.1140/epja/i2003-10144-9} {\bibfield
  {journal} {\bibinfo  {journal} {Eur. Phys. J. A}\ }\textbf {\bibinfo {volume}
  {20}},\ \bibinfo {pages} {293} (\bibinfo {year} {2004})},\ \Eprint
  {http://arxiv.org/abs/nucl-th/0307070} {arXiv:nucl-th/0307070 [nucl-th]}
  \BibitemShut {NoStop}%
\bibitem [{\citenamefont {Navarro~Pérez}\ and\ \citenamefont
  {Lei}(2018)}]{NavarroPerez:2018qzd}%
  \BibitemOpen
  \bibfield  {author} {\bibinfo {author} {\bibfnamefont {R.}~\bibnamefont
  {Navarro~Pérez}}\ and\ \bibinfo {author} {\bibfnamefont {J.}~\bibnamefont
  {Lei}},\ }\href@noop {} {\  (\bibinfo {year} {2018})},\ \Eprint
  {http://arxiv.org/abs/1812.05641} {arXiv:1812.05641 [nucl-th]} \BibitemShut
  {NoStop}%
\bibitem [{\citenamefont {Navarro~Pérez}\ \emph {et~al.}(2014)\citenamefont
  {Navarro~Pérez}, \citenamefont {Amaro},\ and\ \citenamefont
  {Ruiz~Arriola}}]{Perez:2014jsa}%
  \BibitemOpen
  \bibfield  {author} {\bibinfo {author} {\bibfnamefont {R.}~\bibnamefont
  {Navarro~Pérez}}, \bibinfo {author} {\bibfnamefont {J.~E.}\ \bibnamefont
  {Amaro}}, \ and\ \bibinfo {author} {\bibfnamefont {E.}~\bibnamefont
  {Ruiz~Arriola}},\ }\href {\doibase 10.1016/j.physletb.2014.09.035} {\bibfield
   {journal} {\bibinfo  {journal} {Phys. Lett.}\ }\textbf {\bibinfo {volume}
  {B738}},\ \bibinfo {pages} {155} (\bibinfo {year} {2014})},\ \Eprint
  {http://arxiv.org/abs/1407.3937} {arXiv:1407.3937 [nucl-th]} \BibitemShut
  {NoStop}%
\bibitem [{\citenamefont {Nieves}\ and\ \citenamefont
  {Ruiz~Arriola}(2000)}]{Nieves:1999zb}%
  \BibitemOpen
  \bibfield  {author} {\bibinfo {author} {\bibfnamefont {J.}~\bibnamefont
  {Nieves}}\ and\ \bibinfo {author} {\bibfnamefont {E.}~\bibnamefont
  {Ruiz~Arriola}},\ }\href {\doibase 10.1007/s10050-000-4511-0} {\bibfield
  {journal} {\bibinfo  {journal} {Eur. Phys. J.}\ }\textbf {\bibinfo {volume}
  {A8}},\ \bibinfo {pages} {377} (\bibinfo {year} {2000})},\ \Eprint
  {http://arxiv.org/abs/hep-ph/9906437} {arXiv:hep-ph/9906437 [hep-ph]}
  \BibitemShut {NoStop}%
\bibitem [{\citenamefont {Bertsch}\ and\ \citenamefont
  {Bingham}(2017)}]{Bertsch:2017ilb}%
  \BibitemOpen
  \bibfield  {author} {\bibinfo {author} {\bibfnamefont {G.~F.}\ \bibnamefont
  {Bertsch}}\ and\ \bibinfo {author} {\bibfnamefont {D.}~\bibnamefont
  {Bingham}},\ }\href {\doibase 10.1103/PhysRevLett.119.252501} {\bibfield
  {journal} {\bibinfo  {journal} {Phys. Rev. Lett.}\ }\textbf {\bibinfo
  {volume} {119}},\ \bibinfo {pages} {252501} (\bibinfo {year} {2017})},\
  \Eprint {http://arxiv.org/abs/1703.08844} {arXiv:1703.08844 [nucl-th]}
  \BibitemShut {NoStop}%
\bibitem [{\citenamefont {Pastore}(2018)}]{Pastore:2018xuu}%
  \BibitemOpen
  \bibfield  {author} {\bibinfo {author} {\bibfnamefont {A.}~\bibnamefont
  {Pastore}},\ }\href@noop {} {\  (\bibinfo {year} {2018})},\ \Eprint
  {http://arxiv.org/abs/1810.05585} {arXiv:1810.05585 [nucl-th]} \BibitemShut
  {NoStop}%
\bibitem [{\citenamefont {Krupina}\ \emph {et~al.}(2018)\citenamefont
  {Krupina}, \citenamefont {Lensky},\ and\ \citenamefont
  {Pascalutsa}}]{Krupina:2017pgr}%
  \BibitemOpen
  \bibfield  {author} {\bibinfo {author} {\bibfnamefont {N.}~\bibnamefont
  {Krupina}}, \bibinfo {author} {\bibfnamefont {V.}~\bibnamefont {Lensky}}, \
  and\ \bibinfo {author} {\bibfnamefont {V.}~\bibnamefont {Pascalutsa}},\
  }\href {\doibase 10.1016/j.physletb.2018.04.066} {\bibfield  {journal}
  {\bibinfo  {journal} {Phys. Lett. B}\ }\textbf {\bibinfo {volume} {782}},\
  \bibinfo {pages} {34} (\bibinfo {year} {2018})},\ \Eprint
  {http://arxiv.org/abs/1712.05349} {arXiv:1712.05349 [nucl-th]} \BibitemShut
  {NoStop}%
\bibitem [{\citenamefont {Griesshammer}\ \emph {et~al.}(2012)\citenamefont
  {Griesshammer}, \citenamefont {McGovern}, \citenamefont {Phillips},\ and\
  \citenamefont {Feldman}}]{Griesshammer:2012we}%
  \BibitemOpen
  \bibfield  {author} {\bibinfo {author} {\bibfnamefont {H.~W.}\ \bibnamefont
  {Griesshammer}}, \bibinfo {author} {\bibfnamefont {J.~A.}\ \bibnamefont
  {McGovern}}, \bibinfo {author} {\bibfnamefont {D.~R.}\ \bibnamefont
  {Phillips}}, \ and\ \bibinfo {author} {\bibfnamefont {G.}~\bibnamefont
  {Feldman}},\ }\href {\doibase 10.1016/j.ppnp.2012.04.003} {\bibfield
  {journal} {\bibinfo  {journal} {Prog. Part. Nucl. Phys.}\ }\textbf {\bibinfo
  {volume} {67}},\ \bibinfo {pages} {841} (\bibinfo {year} {2012})},\ \Eprint
  {http://arxiv.org/abs/1203.6834} {arXiv:1203.6834 [nucl-th]} \BibitemShut
  {NoStop}%
\bibitem [{\citenamefont {Drechsel}\ \emph {et~al.}(2007)\citenamefont
  {Drechsel}, \citenamefont {Kamalov},\ and\ \citenamefont
  {Tiator}}]{Drechsel:2007if}%
  \BibitemOpen
  \bibfield  {author} {\bibinfo {author} {\bibfnamefont {D.}~\bibnamefont
  {Drechsel}}, \bibinfo {author} {\bibfnamefont {S.~S.}\ \bibnamefont
  {Kamalov}}, \ and\ \bibinfo {author} {\bibfnamefont {L.}~\bibnamefont
  {Tiator}},\ }\href {\doibase 10.1140/epja/i2007-10490-6} {\bibfield
  {journal} {\bibinfo  {journal} {Eur. Phys. J. A}\ }\textbf {\bibinfo {volume}
  {34}},\ \bibinfo {pages} {69} (\bibinfo {year} {2007})},\ \Eprint
  {http://arxiv.org/abs/0710.0306} {arXiv:0710.0306 [nucl-th]} \BibitemShut
  {NoStop}%
\bibitem [{\citenamefont {Davidson}\ and\ \citenamefont
  {Hinkley}(1997)}]{Davidson-Hinkley}%
  \BibitemOpen
  \bibfield  {author} {\bibinfo {author} {\bibfnamefont {A.~C.}\ \bibnamefont
  {Davidson}}\ and\ \bibinfo {author} {\bibfnamefont {D.~V.}\ \bibnamefont
  {Hinkley}},\ }\href@noop {} {\emph {\bibinfo {title} {Bootstrap Methods and
  their Application}}}\ (\bibinfo  {publisher} {Cambridge University Press},\
  \bibinfo {year} {1997})\BibitemShut {NoStop}%
\bibitem [{\citenamefont {Pedroni}\ \emph {et~al.}()\citenamefont {Pedroni},
  \citenamefont {Sconfietti} \emph {et~al.}}]{stat_paper}%
  \BibitemOpen
  \bibfield  {author} {\bibinfo {author} {\bibfnamefont {P.}~\bibnamefont
  {Pedroni}}, \bibinfo {author} {\bibfnamefont {S.}~\bibnamefont {Sconfietti}},
   \emph {et~al.},\ }\href@noop {} {\emph {\bibinfo {title} {in
  preparation}}}\BibitemShut {NoStop}%
\bibitem [{\citenamefont {Hagelstein}\ \emph {et~al.}(2016)\citenamefont
  {Hagelstein}, \citenamefont {Miskimen},\ and\ \citenamefont
  {Pascalutsa}}]{Hagelstein:2015egb}%
  \BibitemOpen
  \bibfield  {author} {\bibinfo {author} {\bibfnamefont {F.}~\bibnamefont
  {Hagelstein}}, \bibinfo {author} {\bibfnamefont {R.}~\bibnamefont
  {Miskimen}}, \ and\ \bibinfo {author} {\bibfnamefont {V.}~\bibnamefont
  {Pascalutsa}},\ }\href {\doibase 10.1016/j.ppnp.2015.12.001} {\bibfield
  {journal} {\bibinfo  {journal} {Prog. Part. Nucl. Phys.}\ }\textbf {\bibinfo
  {volume} {88}},\ \bibinfo {pages} {29} (\bibinfo {year} {2016})},\ \Eprint
  {http://arxiv.org/abs/1512.03765} {arXiv:1512.03765 [nucl-th]} \BibitemShut
  {NoStop}%
\bibitem [{\citenamefont {Olmos~de Leon}\ \emph {et~al.}(2001)\citenamefont
  {Olmos~de Leon} \emph {et~al.}}]{OlmosdeLeon:2001zn}%
  \BibitemOpen
  \bibfield  {author} {\bibinfo {author} {\bibfnamefont {V.}~\bibnamefont
  {Olmos~de Leon}} \emph {et~al.},\ }\href {\doibase 10.1007/s100500170132}
  {\bibfield  {journal} {\bibinfo  {journal} {Eur. Phys. J. A}\ }\textbf
  {\bibinfo {volume} {10}},\ \bibinfo {pages} {207} (\bibinfo {year}
  {2001})}\BibitemShut {NoStop}%
\bibitem [{\citenamefont {Martel}\ \emph {et~al.}(2015)\citenamefont {Martel}
  \emph {et~al.}}]{Martel:2014pba}%
  \BibitemOpen
  \bibfield  {author} {\bibinfo {author} {\bibfnamefont {P.~P.}\ \bibnamefont
  {Martel}} \emph {et~al.} (\bibinfo {collaboration} {A2}),\ }\href {\doibase
  10.1103/PhysRevLett.114.112501} {\bibfield  {journal} {\bibinfo  {journal}
  {Phys. Rev. Lett.}\ }\textbf {\bibinfo {volume} {114}},\ \bibinfo {pages}
  {112501} (\bibinfo {year} {2015})},\ \Eprint {http://arxiv.org/abs/1408.1576}
  {arXiv:1408.1576 [nucl-ex]} \BibitemShut {NoStop}%
\bibitem [{\citenamefont {Ahrens}\ \emph {et~al.}(2001)\citenamefont {Ahrens}
  \emph {et~al.}}]{Ahrens:2001qt}%
  \BibitemOpen
  \bibfield  {author} {\bibinfo {author} {\bibfnamefont {J.}~\bibnamefont
  {Ahrens}} \emph {et~al.} (\bibinfo {collaboration} {GDH, A2}),\ }\href
  {\doibase 10.1103/PhysRevLett.87.022003} {\bibfield  {journal} {\bibinfo
  {journal} {Phys. Rev. Lett.}\ }\textbf {\bibinfo {volume} {87}},\ \bibinfo
  {pages} {022003} (\bibinfo {year} {2001})},\ \Eprint
  {http://arxiv.org/abs/hep-ex/0105089} {arXiv:hep-ex/0105089 [hep-ex]}
  \BibitemShut {NoStop}%
\bibitem [{\citenamefont {Dutz}\ \emph {et~al.}(2003)\citenamefont {Dutz} \emph
  {et~al.}}]{Dutz:2003mm}%
  \BibitemOpen
  \bibfield  {author} {\bibinfo {author} {\bibfnamefont {H.}~\bibnamefont
  {Dutz}} \emph {et~al.} (\bibinfo {collaboration} {GDH}),\ }\href {\doibase
  10.1103/PhysRevLett.91.192001} {\bibfield  {journal} {\bibinfo  {journal}
  {Phys. Rev. Lett.}\ }\textbf {\bibinfo {volume} {91}},\ \bibinfo {pages}
  {192001} (\bibinfo {year} {2003})}\BibitemShut {NoStop}%
\bibitem [{\citenamefont {Pasquini}\ \emph {et~al.}(2010)\citenamefont
  {Pasquini}, \citenamefont {Pedroni},\ and\ \citenamefont
  {Drechsel}}]{Pasquini:2010zr}%
  \BibitemOpen
  \bibfield  {author} {\bibinfo {author} {\bibfnamefont {B.}~\bibnamefont
  {Pasquini}}, \bibinfo {author} {\bibfnamefont {P.}~\bibnamefont {Pedroni}}, \
  and\ \bibinfo {author} {\bibfnamefont {D.}~\bibnamefont {Drechsel}},\ }\href
  {\doibase 10.1016/j.physletb.2010.03.007} {\bibfield  {journal} {\bibinfo
  {journal} {Phys. Lett. B}\ }\textbf {\bibinfo {volume} {687}},\ \bibinfo
  {pages} {160} (\bibinfo {year} {2010})},\ \Eprint
  {http://arxiv.org/abs/1001.4230} {arXiv:1001.4230 [hep-ph]} \BibitemShut
  {NoStop}%
\bibitem [{\citenamefont {Gryniuk}\ \emph {et~al.}(2016)\citenamefont
  {Gryniuk}, \citenamefont {Hagelstein},\ and\ \citenamefont
  {Pascalutsa}}]{Gryniuk:2016gnm}%
  \BibitemOpen
  \bibfield  {author} {\bibinfo {author} {\bibfnamefont {O.}~\bibnamefont
  {Gryniuk}}, \bibinfo {author} {\bibfnamefont {F.}~\bibnamefont {Hagelstein}},
  \ and\ \bibinfo {author} {\bibfnamefont {V.}~\bibnamefont {Pascalutsa}},\
  }\href {\doibase 10.1103/PhysRevD.94.034043} {\bibfield  {journal} {\bibinfo
  {journal} {Phys. Rev.}\ }\textbf {\bibinfo {volume} {D94}},\ \bibinfo {pages}
  {034043} (\bibinfo {year} {2016})},\ \Eprint
  {http://arxiv.org/abs/1604.00789} {arXiv:1604.00789 [nucl-th]} \BibitemShut
  {NoStop}%
\bibitem [{\citenamefont {Wolf}\ \emph {et~al.}(2001)\citenamefont {Wolf} \emph
  {et~al.}}]{Wolf:2001ha}%
  \BibitemOpen
  \bibfield  {author} {\bibinfo {author} {\bibfnamefont {S.}~\bibnamefont
  {Wolf}} \emph {et~al.},\ }\href {\doibase 10.1007/s100500170031} {\bibfield
  {journal} {\bibinfo  {journal} {Eur. Phys. J.}\ }\textbf {\bibinfo {volume}
  {A12}},\ \bibinfo {pages} {231} (\bibinfo {year} {2001})},\ \Eprint
  {http://arxiv.org/abs/nucl-ex/0109013} {arXiv:nucl-ex/0109013 [nucl-ex]}
  \BibitemShut {NoStop}%
\bibitem [{\citenamefont {Camen}\ \emph {et~al.}(2002)\citenamefont {Camen}
  \emph {et~al.}}]{Camen:2001st}%
  \BibitemOpen
  \bibfield  {author} {\bibinfo {author} {\bibfnamefont {M.}~\bibnamefont
  {Camen}} \emph {et~al.},\ }\href {\doibase 10.1103/PhysRevC.65.032202}
  {\bibfield  {journal} {\bibinfo  {journal} {Phys. Rev.}\ }\textbf {\bibinfo
  {volume} {C65}},\ \bibinfo {pages} {032202} (\bibinfo {year} {2002})},\
  \Eprint {http://arxiv.org/abs/nucl-ex/0112015} {arXiv:nucl-ex/0112015
  [nucl-ex]} \BibitemShut {NoStop}%
\bibitem [{\citenamefont {Hinkley}(1977)}]{ref:giacomo}%
  \BibitemOpen
  \bibfield  {author} {\bibinfo {author} {\bibfnamefont {D.~V.}\ \bibnamefont
  {Hinkley}},\ }\href {\doibase 10.2307/2335765} {\bibfield  {journal}
  {\bibinfo  {journal} {Biometrika}\ }\textbf {\bibinfo {volume} {64}},\
  \bibinfo {pages} {21} (\bibinfo {year} {1977})}\BibitemShut {NoStop}%
\bibitem [{\citenamefont {Oxley}(1958)}]{Oxley:1958zz}%
  \BibitemOpen
  \bibfield  {author} {\bibinfo {author} {\bibfnamefont {C.~L.}\ \bibnamefont
  {Oxley}},\ }\href {\doibase 10.1103/PhysRev.110.733} {\bibfield  {journal}
  {\bibinfo  {journal} {Phys. Rev.}\ }\textbf {\bibinfo {volume} {110}},\
  \bibinfo {pages} {733} (\bibinfo {year} {1958})}\BibitemShut {NoStop}%
\bibitem [{\citenamefont {Baranov}\ \emph {et~al.}(1975)\citenamefont
  {Baranov}, \citenamefont {Buinov}, \citenamefont {Godin}, \citenamefont
  {Kuznetsova}, \citenamefont {Petrunkin}, \citenamefont {Tatarinskaya},
  \citenamefont {Shirchenko}, \citenamefont {Shtarkov}, \citenamefont
  {Yurchenko},\ and\ \citenamefont {Yanulis}}]{Baranov:1975ju}%
  \BibitemOpen
  \bibfield  {author} {\bibinfo {author} {\bibfnamefont {P.~S.}\ \bibnamefont
  {Baranov}}, \bibinfo {author} {\bibfnamefont {G.~M.}\ \bibnamefont {Buinov}},
  \bibinfo {author} {\bibfnamefont {V.~G.}\ \bibnamefont {Godin}}, \bibinfo
  {author} {\bibfnamefont {V.~A.}\ \bibnamefont {Kuznetsova}}, \bibinfo
  {author} {\bibfnamefont {V.~A.}\ \bibnamefont {Petrunkin}}, \bibinfo {author}
  {\bibfnamefont {L.~S.}\ \bibnamefont {Tatarinskaya}}, \bibinfo {author}
  {\bibfnamefont {V.~S.}\ \bibnamefont {Shirchenko}}, \bibinfo {author}
  {\bibfnamefont {L.~N.}\ \bibnamefont {Shtarkov}}, \bibinfo {author}
  {\bibfnamefont {V.~V.}\ \bibnamefont {Yurchenko}}, \ and\ \bibinfo {author}
  {\bibfnamefont {{\relax Yu}.~P.}\ \bibnamefont {Yanulis}},\ }\href@noop {}
  {\bibfield  {journal} {\bibinfo  {journal} {Yad. Fiz.}\ }\textbf {\bibinfo
  {volume} {21}},\ \bibinfo {pages} {689} (\bibinfo {year} {1975})}\BibitemShut
  {NoStop}%
\bibitem [{\citenamefont {Birge}(1932)}]{ref:birge}%
  \BibitemOpen
  \bibfield  {author} {\bibinfo {author} {\bibfnamefont {R.~T.}\ \bibnamefont
  {Birge}},\ }\href {\doibase 10.1103/PhysRev.40.207} {\bibfield  {journal}
  {\bibinfo  {journal} {Phys. Rev.}\ }\textbf {\bibinfo {volume} {40}},\
  \bibinfo {pages} {207} (\bibinfo {year} {1932})}\BibitemShut {NoStop}%
\bibitem [{\citenamefont {Behnke}\ \emph {et~al.}(2013)\citenamefont {Behnke},
  \citenamefont {Kroeninger}, \citenamefont {Schott},\ and\ \citenamefont
  {Schoerner-Sadenius~(eds.)}}]{ref:statbook}%
  \BibitemOpen
  \bibfield  {author} {\bibinfo {author} {\bibfnamefont {O.}~\bibnamefont
  {Behnke}}, \bibinfo {author} {\bibfnamefont {K.}~\bibnamefont {Kroeninger}},
  \bibinfo {author} {\bibfnamefont {G.}~\bibnamefont {Schott}}, \ and\ \bibinfo
  {author} {\bibfnamefont {T.}~\bibnamefont {Schoerner-Sadenius~(eds.)}},\
  }\href@noop {} {\emph {\bibinfo {title} {{Data Analysis in High Energy
  Physics: A Practical Guide to Statistical Methods}}}}\ (\bibinfo  {publisher}
  {Wiley-VCH},\ \bibinfo {year} {2013})\BibitemShut {NoStop}%
\bibitem [{\citenamefont {MacGibbon}\ \emph {et~al.}(1995)\citenamefont
  {MacGibbon}, \citenamefont {Garino}, \citenamefont {Lucas}, \citenamefont
  {Nathan}, \citenamefont {Feldman},\ and\ \citenamefont
  {Dolbilkin}}]{MacGibbon:1995in}%
  \BibitemOpen
  \bibfield  {author} {\bibinfo {author} {\bibfnamefont {B.~E.}\ \bibnamefont
  {MacGibbon}}, \bibinfo {author} {\bibfnamefont {G.}~\bibnamefont {Garino}},
  \bibinfo {author} {\bibfnamefont {M.~A.}\ \bibnamefont {Lucas}}, \bibinfo
  {author} {\bibfnamefont {A.~M.}\ \bibnamefont {Nathan}}, \bibinfo {author}
  {\bibfnamefont {G.}~\bibnamefont {Feldman}}, \ and\ \bibinfo {author}
  {\bibfnamefont {B.}~\bibnamefont {Dolbilkin}},\ }\href {\doibase
  10.1103/PhysRevC.52.2097} {\bibfield  {journal} {\bibinfo  {journal} {Phys.
  Rev. C}\ }\textbf {\bibinfo {volume} {52}},\ \bibinfo {pages} {2097}
  (\bibinfo {year} {1995})},\ \Eprint {http://arxiv.org/abs/nucl-ex/9507001}
  {arXiv:nucl-ex/9507001 [nucl-ex]} \BibitemShut {NoStop}%
\bibitem [{\citenamefont {Federspiel}\ \emph {et~al.}(1991)\citenamefont
  {Federspiel}, \citenamefont {Eisenstein}, \citenamefont {Lucas},
  \citenamefont {MacGibbon}, \citenamefont {Mellendorf}, \citenamefont
  {Nathan}, \citenamefont {O'Neill},\ and\ \citenamefont
  {Wells}}]{Federspiel:1991yd}%
  \BibitemOpen
  \bibfield  {author} {\bibinfo {author} {\bibfnamefont {F.~J.}\ \bibnamefont
  {Federspiel}}, \bibinfo {author} {\bibfnamefont {R.~A.}\ \bibnamefont
  {Eisenstein}}, \bibinfo {author} {\bibfnamefont {M.~A.}\ \bibnamefont
  {Lucas}}, \bibinfo {author} {\bibfnamefont {B.~E.}\ \bibnamefont
  {MacGibbon}}, \bibinfo {author} {\bibfnamefont {K.}~\bibnamefont
  {Mellendorf}}, \bibinfo {author} {\bibfnamefont {A.~M.}\ \bibnamefont
  {Nathan}}, \bibinfo {author} {\bibfnamefont {A.}~\bibnamefont {O'Neill}}, \
  and\ \bibinfo {author} {\bibfnamefont {D.~P.}\ \bibnamefont {Wells}},\ }\href
  {\doibase 10.1103/PhysRevLett.67.1511} {\bibfield  {journal} {\bibinfo
  {journal} {Phys. Rev. Lett.}\ }\textbf {\bibinfo {volume} {67}},\ \bibinfo
  {pages} {1511} (\bibinfo {year} {1991})}\BibitemShut {NoStop}%
\bibitem [{\citenamefont {Zieger}\ \emph {et~al.}(1992)\citenamefont {Zieger},
  \citenamefont {Van~de Vyver}, \citenamefont {Christmann}, \citenamefont
  {De~Graeve}, \citenamefont {Van~den Abeele},\ and\ \citenamefont
  {Ziegler}}]{Zieger:1992jq}%
  \BibitemOpen
  \bibfield  {author} {\bibinfo {author} {\bibfnamefont {A.}~\bibnamefont
  {Zieger}}, \bibinfo {author} {\bibfnamefont {R.}~\bibnamefont {Van~de
  Vyver}}, \bibinfo {author} {\bibfnamefont {D.}~\bibnamefont {Christmann}},
  \bibinfo {author} {\bibfnamefont {A.}~\bibnamefont {De~Graeve}}, \bibinfo
  {author} {\bibfnamefont {C.}~\bibnamefont {Van~den Abeele}}, \ and\ \bibinfo
  {author} {\bibfnamefont {B.}~\bibnamefont {Ziegler}},\ }\href {\doibase
  10.1016/0370-2693(92)90707-B} {\bibfield  {journal} {\bibinfo  {journal}
  {Phys. Lett. B}\ }\textbf {\bibinfo {volume} {278}},\ \bibinfo {pages} {34}
  (\bibinfo {year} {1992})}\BibitemShut {NoStop}%
\bibitem [{\citenamefont {Griesshammer}\ \emph {et~al.}(2016)\citenamefont
  {Griesshammer}, \citenamefont {McGovern},\ and\ \citenamefont
  {Phillips}}]{Griesshammer:2015ahu}%
  \BibitemOpen
  \bibfield  {author} {\bibinfo {author} {\bibfnamefont {H.~W.}\ \bibnamefont
  {Griesshammer}}, \bibinfo {author} {\bibfnamefont {J.~A.}\ \bibnamefont
  {McGovern}}, \ and\ \bibinfo {author} {\bibfnamefont {D.~R.}\ \bibnamefont
  {Phillips}},\ }\href {\doibase 10.1140/epja/i2016-16139-5} {\bibfield
  {journal} {\bibinfo  {journal} {Eur. Phys. J.}\ }\textbf {\bibinfo {volume}
  {A52}},\ \bibinfo {pages} {139} (\bibinfo {year} {2016})},\ \Eprint
  {http://arxiv.org/abs/1511.01952} {arXiv:1511.01952 [nucl-th]} \BibitemShut
  {NoStop}%
\bibitem [{\citenamefont {Lensky}\ and\ \citenamefont
  {McGovern}(2014)}]{Lensky:2014efa}%
  \BibitemOpen
  \bibfield  {author} {\bibinfo {author} {\bibfnamefont {V.}~\bibnamefont
  {Lensky}}\ and\ \bibinfo {author} {\bibfnamefont {J.~A.}\ \bibnamefont
  {McGovern}},\ }\href {\doibase 10.1103/PhysRevC.89.032202} {\bibfield
  {journal} {\bibinfo  {journal} {Phys. Rev.}\ }\textbf {\bibinfo {volume}
  {C89}},\ \bibinfo {pages} {032202} (\bibinfo {year} {2014})},\ \Eprint
  {http://arxiv.org/abs/1401.3320} {arXiv:1401.3320 [nucl-th]} \BibitemShut
  {NoStop}%
\bibitem [{\citenamefont {Patrignani}\ \emph {et~al.}(2016)\citenamefont
  {Patrignani} \emph {et~al.}}]{Patrignani:2016xqp}%
  \BibitemOpen
  \bibfield  {author} {\bibinfo {author} {\bibfnamefont {C.}~\bibnamefont
  {Patrignani}} \emph {et~al.} (\bibinfo {collaboration} {Particle Data
  Group}),\ }\href {\doibase 10.1088/1674-1137/40/10/100001} {\bibfield
  {journal} {\bibinfo  {journal} {Chin. Phys. C}\ }\textbf {\bibinfo {volume}
  {40}},\ \bibinfo {pages} {100001} (\bibinfo {year} {2016})}\BibitemShut
  {NoStop}%
\bibitem [{\citenamefont {Gryniuk}\ \emph {et~al.}(2015)\citenamefont
  {Gryniuk}, \citenamefont {Hagelstein},\ and\ \citenamefont
  {Pascalutsa}}]{Gryniuk:2015eza}%
  \BibitemOpen
  \bibfield  {author} {\bibinfo {author} {\bibfnamefont {O.}~\bibnamefont
  {Gryniuk}}, \bibinfo {author} {\bibfnamefont {F.}~\bibnamefont {Hagelstein}},
  \ and\ \bibinfo {author} {\bibfnamefont {V.}~\bibnamefont {Pascalutsa}},\
  }\href {\doibase 10.1103/PhysRevD.92.074031} {\bibfield  {journal} {\bibinfo
  {journal} {Phys. Rev.}\ }\textbf {\bibinfo {volume} {D92}},\ \bibinfo {pages}
  {074031} (\bibinfo {year} {2015})},\ \Eprint
  {http://arxiv.org/abs/1508.07952} {arXiv:1508.07952 [nucl-th]} \BibitemShut
  {NoStop}%
\bibitem [{\citenamefont {Sokhoyan}\ \emph {et~al.}(2017)\citenamefont
  {Sokhoyan} \emph {et~al.}}]{A2-epja}%
  \BibitemOpen
  \bibfield  {author} {\bibinfo {author} {\bibfnamefont {V.}~\bibnamefont
  {Sokhoyan}} \emph {et~al.},\ }\href {\doibase 10.1140/epja/i2017-12203-0}
  {\bibfield  {journal} {\bibinfo  {journal} {Eur. Phys. J.}\ }\textbf
  {\bibinfo {volume} {A53}},\ \bibinfo {pages} {14} (\bibinfo {year} {2017})},\
  \Eprint {http://arxiv.org/abs/1611.03769} {arXiv:1611.03769 [nucl-ex]}
  \BibitemShut {NoStop}%
\bibitem [{\citenamefont {Downie}\ and\ \citenamefont
  {et~al.}(2016)}]{A2-Downie}%
  \BibitemOpen
  \bibfield  {author} {\bibinfo {author} {\bibfnamefont {E.~J.}\ \bibnamefont
  {Downie}}\ and\ \bibinfo {author} {\bibnamefont {et~al.}},\ }\href@noop {}
  {\bibfield  {journal} {\bibinfo  {journal} {Proposal MAMI-A2/04-16}\ }
  (\bibinfo {year} {2016})}\BibitemShut {NoStop}%
\bibitem [{\citenamefont {Hyman}\ \emph {et~al.}(1959)\citenamefont {Hyman},
  \citenamefont {Ely}, \citenamefont {Frisch},\ and\ \citenamefont
  {Wahlig}}]{Hyman:1959zz}%
  \BibitemOpen
  \bibfield  {author} {\bibinfo {author} {\bibfnamefont {L.~G.}\ \bibnamefont
  {Hyman}}, \bibinfo {author} {\bibfnamefont {R.}~\bibnamefont {Ely}}, \bibinfo
  {author} {\bibfnamefont {D.~H.}\ \bibnamefont {Frisch}}, \ and\ \bibinfo
  {author} {\bibfnamefont {M.~A.}\ \bibnamefont {Wahlig}},\ }\href {\doibase
  10.1103/PhysRevLett.3.93} {\bibfield  {journal} {\bibinfo  {journal} {Phys.
  Rev. Lett.}\ }\textbf {\bibinfo {volume} {3}},\ \bibinfo {pages} {93}
  (\bibinfo {year} {1959})}\BibitemShut {NoStop}%
\bibitem [{\citenamefont {Goldansky}\ \emph {et~al.}(1960)\citenamefont
  {Goldansky}, \citenamefont {Karpukhin}, \citenamefont {Kutsenko},\ and\
  \citenamefont {Pavlovskaya}}]{GOLDANSKY1960473}%
  \BibitemOpen
  \bibfield  {author} {\bibinfo {author} {\bibfnamefont {V.}~\bibnamefont
  {Goldansky}}, \bibinfo {author} {\bibfnamefont {O.}~\bibnamefont
  {Karpukhin}}, \bibinfo {author} {\bibfnamefont {A.}~\bibnamefont {Kutsenko}},
  \ and\ \bibinfo {author} {\bibfnamefont {V.}~\bibnamefont {Pavlovskaya}},\
  }\href {\doibase https://doi.org/10.1016/0029-5582(60)90418-1} {\bibfield
  {journal} {\bibinfo  {journal} {Nuclear Physics}\ }\textbf {\bibinfo {volume}
  {18}},\ \bibinfo {pages} {473 } (\bibinfo {year} {1960})}\BibitemShut
  {NoStop}%
\bibitem [{\citenamefont {Pugh}\ \emph {et~al.}(1957)\citenamefont {Pugh},
  \citenamefont {Gomez}, \citenamefont {Frisch},\ and\ \citenamefont
  {Janes}}]{Pugh:1957zz}%
  \BibitemOpen
  \bibfield  {author} {\bibinfo {author} {\bibfnamefont {G.~E.}\ \bibnamefont
  {Pugh}}, \bibinfo {author} {\bibfnamefont {R.}~\bibnamefont {Gomez}},
  \bibinfo {author} {\bibfnamefont {D.~H.}\ \bibnamefont {Frisch}}, \ and\
  \bibinfo {author} {\bibfnamefont {G.~S.}\ \bibnamefont {Janes}},\ }\href
  {\doibase 10.1103/PhysRev.105.982} {\bibfield  {journal} {\bibinfo  {journal}
  {Phys. Rev.}\ }\textbf {\bibinfo {volume} {105}},\ \bibinfo {pages} {982}
  (\bibinfo {year} {1957})}\BibitemShut {NoStop}%
\bibitem [{\citenamefont {Baranov}\ \emph {et~al.}(2001)\citenamefont
  {Baranov}, \citenamefont {L'vov}, \citenamefont {Petrunkin},\ and\
  \citenamefont {Shtarkov}}]{Baranov:2001jv}%
  \BibitemOpen
  \bibfield  {author} {\bibinfo {author} {\bibfnamefont {P.~S.}\ \bibnamefont
  {Baranov}}, \bibinfo {author} {\bibfnamefont {A.~I.}\ \bibnamefont {L'vov}},
  \bibinfo {author} {\bibfnamefont {V.~A.}\ \bibnamefont {Petrunkin}}, \ and\
  \bibinfo {author} {\bibfnamefont {L.~N.}\ \bibnamefont {Shtarkov}},\
  }\href@noop {} {\bibfield  {journal} {\bibinfo  {journal} {Phys. Part.
  Nucl.}\ }\textbf {\bibinfo {volume} {32}},\ \bibinfo {pages} {376} (\bibinfo
  {year} {2001})},\ \bibinfo {note} {[Fiz. Elem. Chast. Atom.
  Yadra32,699(2001)]}\BibitemShut {NoStop}%
\bibitem [{\citenamefont {Hallin}\ \emph {et~al.}(1993)\citenamefont {Hallin}
  \emph {et~al.}}]{Hallin:1993ft}%
  \BibitemOpen
  \bibfield  {author} {\bibinfo {author} {\bibfnamefont {E.~L.}\ \bibnamefont
  {Hallin}} \emph {et~al.},\ }\href {\doibase 10.1103/PhysRevC.48.1497}
  {\bibfield  {journal} {\bibinfo  {journal} {Phys. Rev. C}\ }\textbf {\bibinfo
  {volume} {48}},\ \bibinfo {pages} {1497} (\bibinfo {year}
  {1993})}\BibitemShut {NoStop}%
\bibitem [{\citenamefont {Bernardini}\ \emph {et~al.}(1960)\citenamefont
  {Bernardini}, \citenamefont {Hanson}, \citenamefont {Odian}, \citenamefont
  {Yamagata}, \citenamefont {Auerbach},\ and\ \citenamefont
  {Filosofo}}]{Bernardini:1960wya}%
  \BibitemOpen
  \bibfield  {author} {\bibinfo {author} {\bibfnamefont {G.}~\bibnamefont
  {Bernardini}}, \bibinfo {author} {\bibfnamefont {A.~O.}\ \bibnamefont
  {Hanson}}, \bibinfo {author} {\bibfnamefont {A.~C.}\ \bibnamefont {Odian}},
  \bibinfo {author} {\bibfnamefont {T.}~\bibnamefont {Yamagata}}, \bibinfo
  {author} {\bibfnamefont {L.~B.}\ \bibnamefont {Auerbach}}, \ and\ \bibinfo
  {author} {\bibfnamefont {I.}~\bibnamefont {Filosofo}},\ }\href {\doibase
  10.1007/BF02733177} {\bibfield  {journal} {\bibinfo  {journal} {Nuovo cim.}\
  }\textbf {\bibinfo {volume} {18}},\ \bibinfo {pages} {1203} (\bibinfo {year}
  {1960})}\BibitemShut {NoStop}%
\bibitem [{\citenamefont {Baranov}\ \emph {et~al.}(1974)\citenamefont
  {Baranov}, \citenamefont {Buinov}, \citenamefont {Godin}, \citenamefont
  {Kuznetzova}, \citenamefont {Petrunkin}, \citenamefont {Tatarinskaya},
  \citenamefont {Shirthenko}, \citenamefont {Shtarkov}, \citenamefont
  {Yurtchenko},\ and\ \citenamefont {Yanulis}}]{Baranov:1974ec}%
  \BibitemOpen
  \bibfield  {author} {\bibinfo {author} {\bibfnamefont {P.}~\bibnamefont
  {Baranov}}, \bibinfo {author} {\bibfnamefont {G.}~\bibnamefont {Buinov}},
  \bibinfo {author} {\bibfnamefont {V.}~\bibnamefont {Godin}}, \bibinfo
  {author} {\bibfnamefont {V.}~\bibnamefont {Kuznetzova}}, \bibinfo {author}
  {\bibfnamefont {V.}~\bibnamefont {Petrunkin}}, \bibinfo {author}
  {\bibnamefont {Tatarinskaya}}, \bibinfo {author} {\bibfnamefont
  {V.}~\bibnamefont {Shirthenko}}, \bibinfo {author} {\bibfnamefont
  {L.}~\bibnamefont {Shtarkov}}, \bibinfo {author} {\bibfnamefont
  {V.}~\bibnamefont {Yurtchenko}}, \ and\ \bibinfo {author} {\bibfnamefont
  {{\relax Yu}.}~\bibnamefont {Yanulis}},\ }\href {\doibase
  10.1016/0370-2693(74)90736-9} {\bibfield  {journal} {\bibinfo  {journal}
  {Phys. Lett.}\ }\textbf {\bibinfo {volume} {52B}},\ \bibinfo {pages} {122}
  (\bibinfo {year} {1974})}\BibitemShut {NoStop}%
\end{thebibliography}
